\documentclass[superscriptaddress,nofootinbib,amsmath,amssymb,aps,prb,twocolumn]{revtex4-2}

\usepackage{graphicx} 
\usepackage{dcolumn}          
\usepackage{bm}               
\usepackage{upgreek}
\usepackage{booktabs} 				
\usepackage{tabularx}
\usepackage{array}
\usepackage[normalem]{ulem}
\usepackage[colorlinks=true,urlcolor=blue,breaklinks=true]{hyperref}

\usepackage{xcolor}
\usepackage{xr}
\usepackage{epstopdf}

\newcommand{\ts}[1]{\textcolor{black}{#1}}
\newcommand{\pz}[1]{\textcolor{black}{#1}}
\newcommand{\rd}[1]{\textcolor{black}{#1}}

\begin{document}

\title{Microwave spectroscopy of Andreev states in InAs nanowire-based hybrid junctions using a flip-chip layout}

\author{Patrick Zellekens$^+$}
\email{p.zellekens@fz-juelich.de}
\altaffiliation{Present address: RIKEN Center for Emergent Matter Science and Advanced Device Laboratory, 351-0198 Saitama, Japan}
\affiliation{Peter Gr\"unberg Institut (PGI-9), Forschungszentrum J\"ulich, 52425 J\"ulich, Germany}
\affiliation{JARA-Fundamentals of Future Information Technology, J\"ulich-Aachen Research Alliance, Forschungszentrum J\"ulich and RWTH Aachen University, Germany}

\author{Russell Deacon$^+$}
\email{russell@riken.jp}
\affiliation{RIKEN Center for Emergent Matter Science and Advanced Device Laboratory, 351-0198 Saitama, Japan}


\author{Pujitha Perla}
\affiliation{Peter Gr\"unberg Institut (PGI-9), Forschungszentrum J\"ulich, 52425 J\"ulich, Germany}
\affiliation{JARA-Fundamentals of Future Information Technology, J\"ulich-Aachen Research Alliance, Forschungszentrum J\"ulich and RWTH Aachen University, Germany}


\author{Detlev Gr\"utzmacher}
\affiliation{Peter Gr\"unberg Institut (PGI-9), Forschungszentrum J\"ulich, 52425 J\"ulich, Germany}
\affiliation{JARA-Fundamentals of Future Information Technology, J\"ulich-Aachen Research Alliance, Forschungszentrum J\"ulich and RWTH Aachen University, Germany}

\author{Mihail Ion Lepsa}
\affiliation{Peter Gr\"unberg Institut (PGI-9), Forschungszentrum J\"ulich, 52425 J\"ulich, Germany}
\affiliation{JARA-Fundamentals of Future Information Technology, J\"ulich-Aachen Research Alliance, Forschungszentrum J\"ulich and RWTH Aachen University, Germany}

\author{Thomas Sch\"apers}
\affiliation{Peter Gr\"unberg Institut (PGI-9), Forschungszentrum J\"ulich, 52425 J\"ulich, Germany}
\affiliation{JARA-Fundamentals of Future Information Technology, J\"ulich-Aachen Research Alliance, Forschungszentrum J\"ulich and RWTH Aachen University, Germany}

\author{Koji Ishibashi}
\affiliation{RIKEN Center for Emergent Matter Science and Advanced Device Laboratory, 351-0198 Saitama, Japan}

\hyphenation{}
\date{\today}

\begin{abstract}
Josephson junctions based on semiconductor nanowires are potential building blocks for electrically tunable qubit structures, e.g. the gatemon or the Andreev qubit. However, an actual realization requires the thorough investigation of the intrinsic excitation spectrum. Here, we demonstrate the fabrication of low-loss superconducting microwave circuits that combine high quality factors with a well-controlled gate architecture by utilizing a flip-chip approach. This platform is then used to perform single-tone and two-tone experiments on Andreev states in in-situ grown InAs/Al core/half-shell nanowires with shadow mask defined Josephson junctions. In gate-controlled and flux-biased spectroscopic measurements we find clear signatures of single quasiparticle as well as quasiparticle pair transitions between discrete Andreev bound states mediated by photon-absorption. Our experimental findings are supported by simulations that show that the junction resides in the intermediate channel length regime. 
\end{abstract}

\maketitle
\def\thefootnote{+}\footnotetext{These authors contributed equally to this work}\def\thefootnote{\arabic{footnote}}



\pz{Over the course of the last decade, Josephson junctions based on superconductor-insulator/normal conductor-superconductor stacks have been used as building blocks for various kinds of quantum bit structures \cite{Clarke08} such as the fluxonium, the transmon, or the gatemon as its electrically-tunable counterpart. All these structures, including the gatemon, have in common that they rely on the collective excitation of an extended circuit structure and only utilize macroscopic properties such as the supercurrent to manipulate and interact with the system.} 

\rd{In contrast to conventional Josephson circuits there are mesosopic systems, which have been already realized by means of different superconductor hybrid structures, including break junctions, carbon nanotubes, and semiconductor nanowires. Here, the characteristics of the system can be determined by a small number of well resolved microscopic excitations that are spatially localized within the normal conducting section, so-called Andreev bound states \cite{Kulik69,Beenakker92a}.} \pz{Transitions between a pair of such states can be utilized in two special types of quantum bits, i.e. the Andreev level qubit \cite{Zazunov03,Zazunov05,Hays18,Javier21} and the Andreev spin qubit \cite{Nazarov03,Nazarov10,Park17,Hays20,Hays21}. In addition study of the energy-phase dispersion of Andreev bound states can provide insight into details of the superconducting state of the system. This is of interest as the range of hybrid systems being studies is expanding from metal and semiconductor based systems to new topological insulator or two-dimensional systems, partly motivated by proposals for novel superconducting states such as proximity induced topological superconductivity \cite{Prada20,HasanKane10}.}

\begin{figure*}[!t]
   \centering
\includegraphics[width=0.8\textwidth]{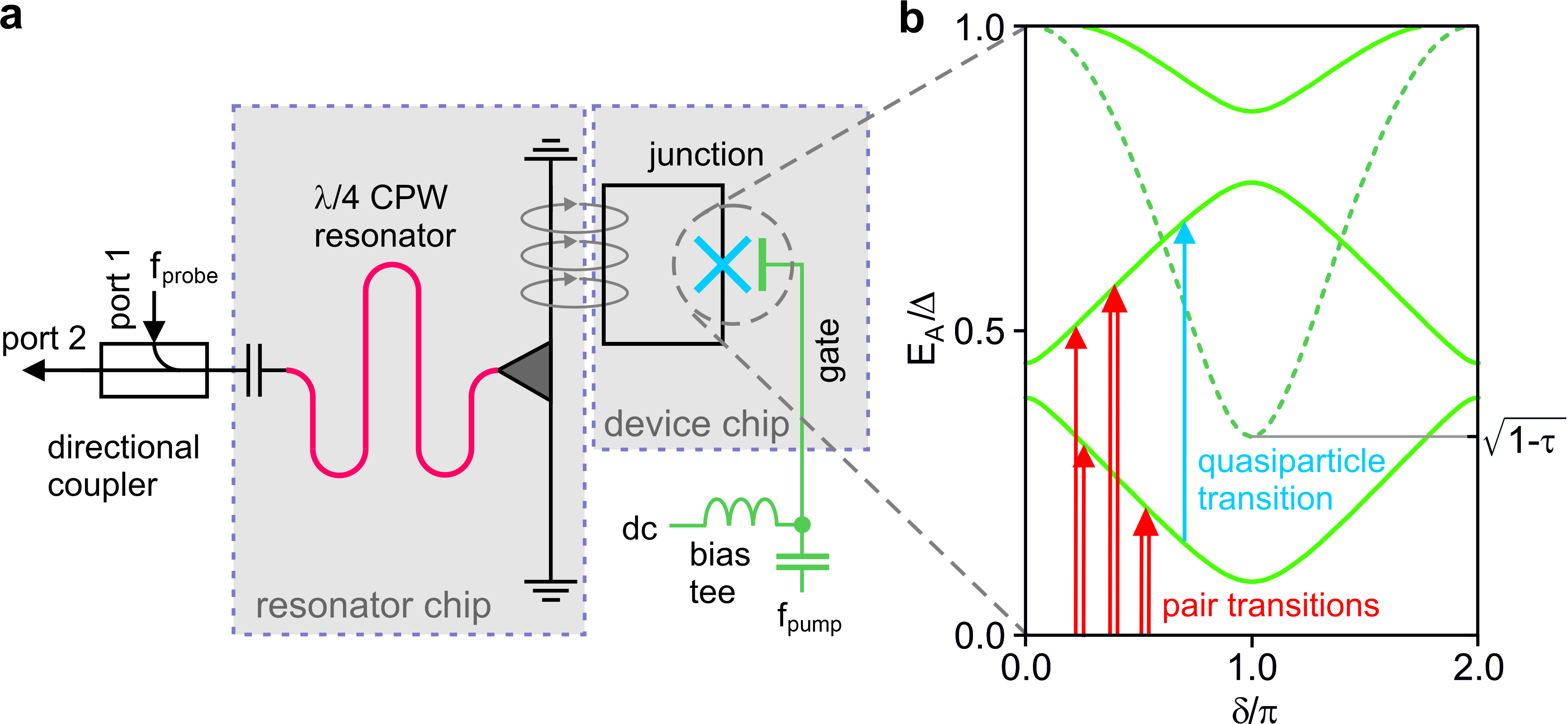}
   \caption{\textbf{Schematic sketch of the measurement setup.} \textbf{a}, The nanowire Josephson junction on the device chip is inductively coupled to the $\lambda/4$ coplanar waveguide (CPW) resonator on the sapphire-based superconducting microwave circuit. The gate is connected to an external bias tee which supplies the dc-bias and the rf signal of the second tone. \textbf{b}, Excitation representation of Andreev levels for a short junction with $L/\xi< 1$ (dashed line) and a long junction with $L/\xi=2.74$ (solid lines). The transparency was set to typical state-of-the-art values for epitaxial nanowire Josephson junctions ($\tau=0.895$). Here, $L$ corresponds to the effective junction length, $\xi=\hbar v_\mathrm{F}/\Delta$ is the superconducting coherence length, $v_\mathrm{F}$ the Fermi velocity and $\Delta$ the superconducting gap. The blue and red arrows indicate possible single quasiparticle (odd) and quasiparticle pair (even) transitions, respectively.}
   \label{fig:setup}
\end{figure*}

\pz{The successful realization of any system based on Andreev bound states requires a deep and detailed knowledge of the energy-phase dispersion of the Andreev levels. In the past, tunnelling measurements have been employed for the spectroscopic analysis of single Andreev bound states \cite{Deacon10,Pillet10}. As an alternative technique, the Andreev bound state spectrum can be detected through coupling of a phase biased junction to a microwave cavity, e.g. a coplanar waveguide (CPW) resonator. This two-tone spectroscopy may provide evidence of the properties of sought after exotic phenomenon such as topologically protected Majorana modes \cite{Peng16,vanHeck17,Prada20}. The coupling of the device with the cavity can be readout using the dispersive shift of the resonance single-tone response and transitions between Andreev bound states can be mapped using two-tone spectroscopy techniques typically employed in circuit quantum electrodynamics \cite{Janvier15,Hays18,Tosi19}.}

\pz{Each new hybrid system has its own set of material processing requirements, e.g. substrate limitations, the sensitivity to chemicals and etchants or the thermal budget. In particular, semiconductor nanowire based devices can require special etching steps and the deposition of high-k gate dielectrics that are known to be counterproductive to the performance of superconducting microwave resonators. Additionally, state-of-the-art fabrication techniques of superconducting microwave circuits such as deep trench etching are usually highly detrimental for semiconductor devices \cite{Bruno15}. Here, we propose a flip-chip method, with the aim of separating the fabrication of the Josephson junction circuit and the microwave circuit. Regarding the latter, the key is the reduction of parasitic two-level systems by optimizing interfaces and surfaces \cite{Gao07,Goeppl08,Barends09,Megrant12,Bruno15} and limiting the effective area on the microwave circuit that is exposed to the dielectric. In addition, one can select low-loss substrate materials such as sapphire to improve the intrinsic quality of the superconducting layer without the risk of a degradation or destruction of the nanowire Josephson junction due to electrostatic discharge. Our flip-chip approach creates a platform for the coupling of superconducting circuits \ts{containing} high Q-factor coplanar waveguide resonators to various hybrid devices, including selectively-grown structures such as topological insulators \cite{Schueffelgen19b}, which can then be used to perform spectroscopy of Andreev states in any highly transparent system. In order to demonstrate the feasibility and performance of our flip-chip approach, we investigated the coupling of the resonator to the device chip. Furthermore, we present single-tone as well as two-tone spectroscopic measurements of an epitaxially grown InAs/Al half shell nanowire with a shadow mask defined Josephson junction \cite{Zellekens20a,Perla21}.} \rd{We note that a similar flip-chip method has recently been reported for Josephson junction detector based spectrometers \cite{Griesmar21}.}
\begin{figure*}[!t]
   \centering
\includegraphics[width=0.95\textwidth]{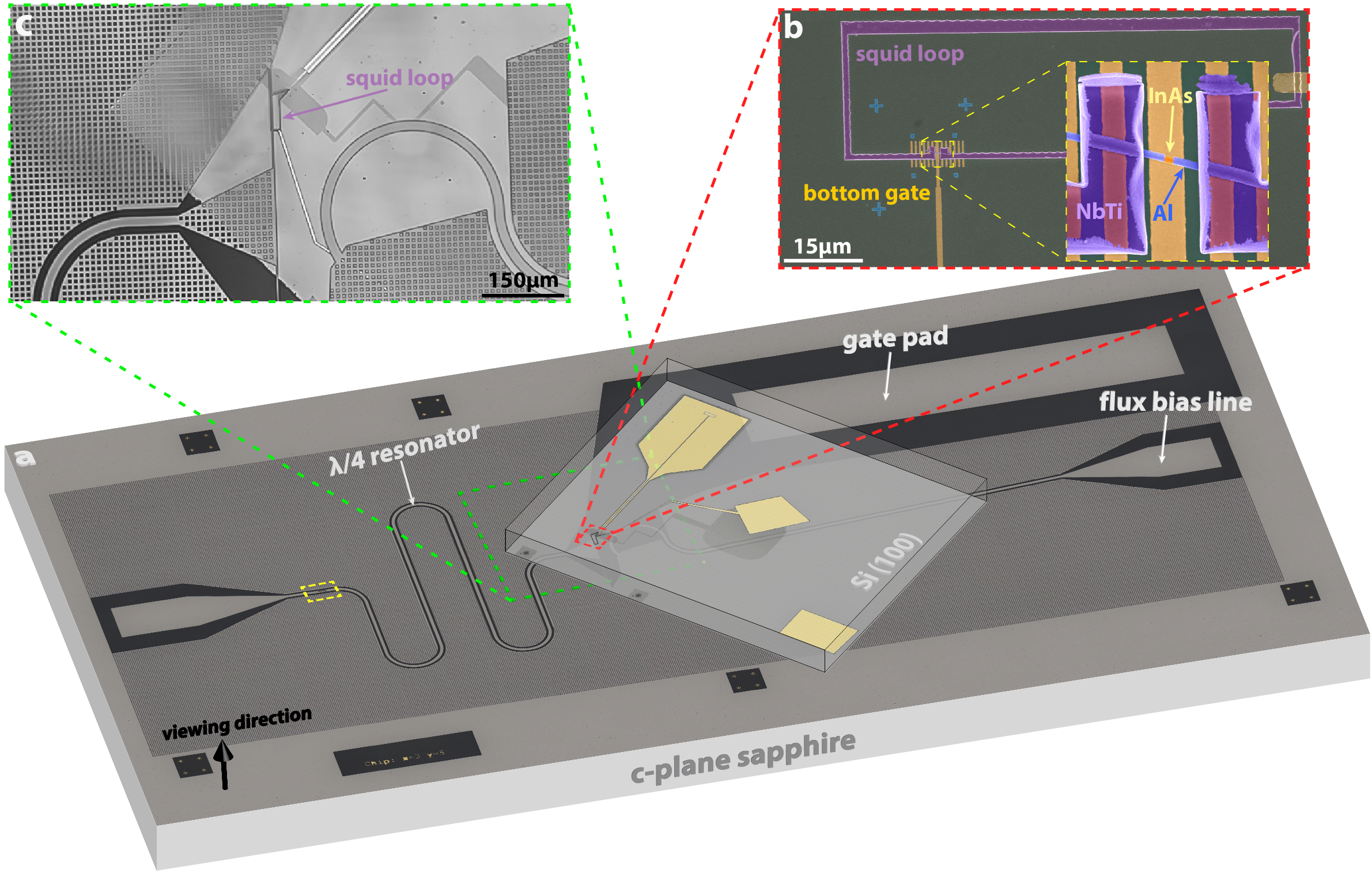}
   \caption{\textbf{Flip-chip layout.} \textbf{a,} Optical micrograph of an exemplary microwave circuit with a $650\,\mu$m-thick sapphire wafer substrate in $c$-plane orientation. All superconducting structures are made out of 50-nm-thick molecular beam epitaxy grown Nb, which has been patterned by means of dry etching. For the application of the microwave signal, a single $\lambda/4$ coplanar waveguide resonator is used, which is capacitively coupled (yellow square) to a short transmission line and an impedance-matched superconducting bonding pad. The electromagnetic tuning of the device is realized with a combination of an on-chip flux bias line and a large bonding pad that acts as the mechanical and electrical connection between the gate electrode on the device chip and the fridge wiring. \textbf{b,} Scanning electron microscopy image of an in-situ formed InAs/Al nanowire Josephson junction on top of a single bottom gate electrode. The SQUID loop itself and the electrical contacts are made out of 80-nm-thick NbTi. \textbf{c,} Optical micrograph of a flipped device. The image was taken through the sapphire substrate mounted upon a polymer stamp during alignment. The SQUID loop is aligned such that one side is located above the termination point of the resonator. To ensure a constant and homogeneous reference potential, the device is connected to a large bonding pad that is electrically coupled to the ground plane of the sapphire chip with a conducting epoxy.}
   \label{fig:Sapphire}
\end{figure*}

\begin{figure*}[!t]
   \centering
\includegraphics[width=0.99\textwidth]{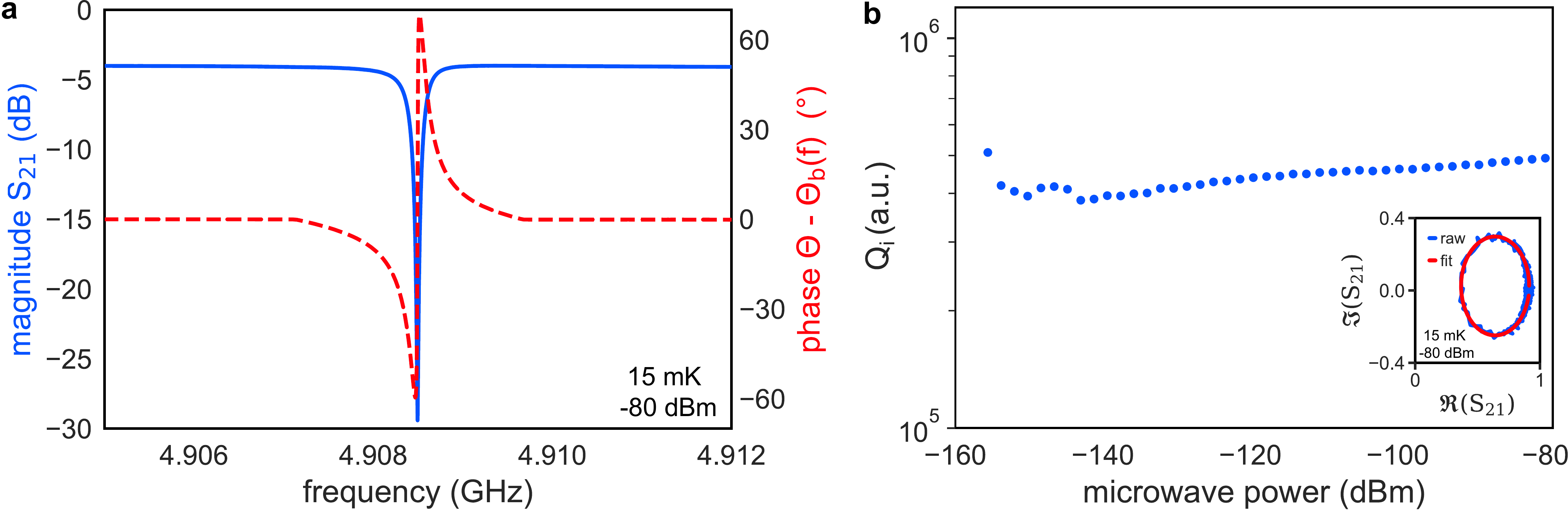}
   \caption{\textbf{Resonator characterization.} \textbf{a,} Single-tone response of a strongly overcoupled resonator similar to the microwave circuit shown in \textbf{a} with a sample chip attached to it and with the nanowire Josephson junction set into pinch-off. The blue line corresponds to the measured change in magnitude and the red line to the unwrapped phase, respectively. The latter has been corrected by the frequency-dependent background phase $\Theta_b(f)$. \textbf{b,} Extracted internal quality factor of the very same resonator as a function of the applied microwave power. Measurements are performed at $T\sim 15\,$mK and the rf-power is indicated with the fridge-internal attenuation accounted for. The inset shows an exemplary circle fit of the resonator in the IQ plane at an applied microwave power of $-80\,$dBm.}
   \label{fig:resonance}
\end{figure*}

\section*{Flip-chip technique}


\pz{A schematic representation of our system, consisting of the microwave circuit on sapphire (resonator chip) and the Josephson junction device on high-resistivity silicon (100) (device chip), is shown in Fig.~\ref{fig:setup}a. Here, a rf-SQUID, comprised of a gate-controlled nanowire Josephson junction in a superconducting loop, is inductively coupled to a quarter wavelength ($\lambda/4$) CPW resonator. By this approach, it is possible to make use of well-defined pre-patterned surface gates and high-k dielectrics with low defect density on the device chip and still preserve high quality microwave structures. The superconducting Nb resonator is connected to a directional coupler, thus providing the pump tone as the input signal and feeding the reflected signal back to the measurement setup. The mesoscopic nanowire Josephson junction is based on an epitaxial Al-InAs half-shell nanowire Josephson junction device that has been prepared by utilizing an in-situ shadow evaporation process \cite{Zellekens20a}. In Fig.~\ref{fig:Sapphire} an optical micrograph of the microwave circuit together with details of the alignment of the device chip on the resonator circuit as well as a micrograph of the nanowire-based Josephson junction are depicted. Further details about the mounting procedure and an image of a successfully mounted chip are provided in the Supplementary Information.}

\pz{When coupled to the microwave cavity, the resonator photons drive transitions between Andreev levels and with sufficient coupling the state of the system is detected via the dispersive shift of the cavity resonance \cite{Janvier15}, i.e. the single-tone response. Here, two types of excitations of the Andreev system are possible. First, the microwave photon assisted excitation of a pair from the superconducting condensate into the Andreev doublet with total energy $(E_{A,1}+E_{A,2})$, where $E_{A,n}$ is the energy of Andreev state $n$. Second, the excitation of a single quasiparticle from one Andreev level to another \cite{Park17,Tosi19}. Both transitions are in literature typically identified by the parity of the number of quasi-particles involved, i.e. even (pair) or odd (single). An example Andreev bound state spectrum in the excitation representation is shown in Fig.~\ref{fig:setup}b where each line indicates a doublet of spin degenerate states. \ts{The details of the energy dispersion depend not just on the superconducting gap and junction length but also properties of the weak link system such as spin-orbit interaction or confinement potential. For longer channels with $L/\xi>1$, higher Andreev bound states are folded into the phase space, as indicated by solid lines Fig.~\ref{fig:setup}b. The dashed line in Fig.~\ref{fig:setup}b corresponds to the single channel/short junction limit, which has been utilized already to study Rabi oscillations in both atomic break junctions \cite{Janvier15} and semiconductor nanowire-based junctions \cite{Hays18}.} In the limit of a short junction, with length $L$ much smaller than the superconducting coherence length $\xi$, a doublet of Andreev bound states appear at energies $E_\mathrm{A}=\Delta \sqrt{1-\tau\sin^2{(\varphi/2)}}$, where $\tau$ is the junction transparency and $\varphi$ the superconducting phase difference \cite{Beenakker91,Furusaki91,Furusaki99}.} 
 



\section*{Determination of resonator properties}

\ts{As a first step to characterize the flip-chip set-up, the properties of the $\lambda/4$ CPW resonators are determined.} Figure~\ref{fig:resonance}a provides an example for the single-tone response $S_{21}$ of a highly overcoupled ($Q_{\text{c}}\approx 20000$) resonator similar to the one shown in Fig.~\ref{fig:Sapphire} with the nanowire device gated such that all states are detuned from the cavity and measured in a \ts{dilution} refrigerator at $T\sim 15\,$mK. For an applied microwave power of -80$\,$dBm, a narrow dip in magnitude (solid blue line) can be observed \ts{at resonance}. In addition, the resonator exhibits a pronounced phase shift  \ts{at this frequency (dashed red line).} Here, the phase signal has been corrected for the monotonically changing, frequency-dependent background phase $\Theta_b$. Figure~\ref{fig:resonance}b shows the power-dependent  internal quality factor $Q_i$ of the very same chip as in Fig.~\ref{fig:resonance}a. The microwave power is estimated at the resonator chip including $62\,$dB of cold attenuation inside the dilution fridge  and external attenuation. The chip reveals a maximum of \ts{$Q_{\text{i}}=5\times 10^5$} in transmission-mode $S_{21}$. Based on an estimated single photon limit of $\sim 150\,$dBm, our resonator reveals a large internal $Q$ factor even at low photon numbers with the decrease in $Q_{i}$ anticipated to arise from the increasing influence of unsaturated two-level systems in the device. The characteristic parameters of the resonator are thereby determined by means of the circle fit method \cite{Probst15} (see inset in Figure~\ref{fig:resonance}b and the Supplementary Information for further details). 

\section*{Interaction of cavity and short channel junction}

\begin{figure*}[!t]
   \centering
\includegraphics[width=0.70\textwidth]{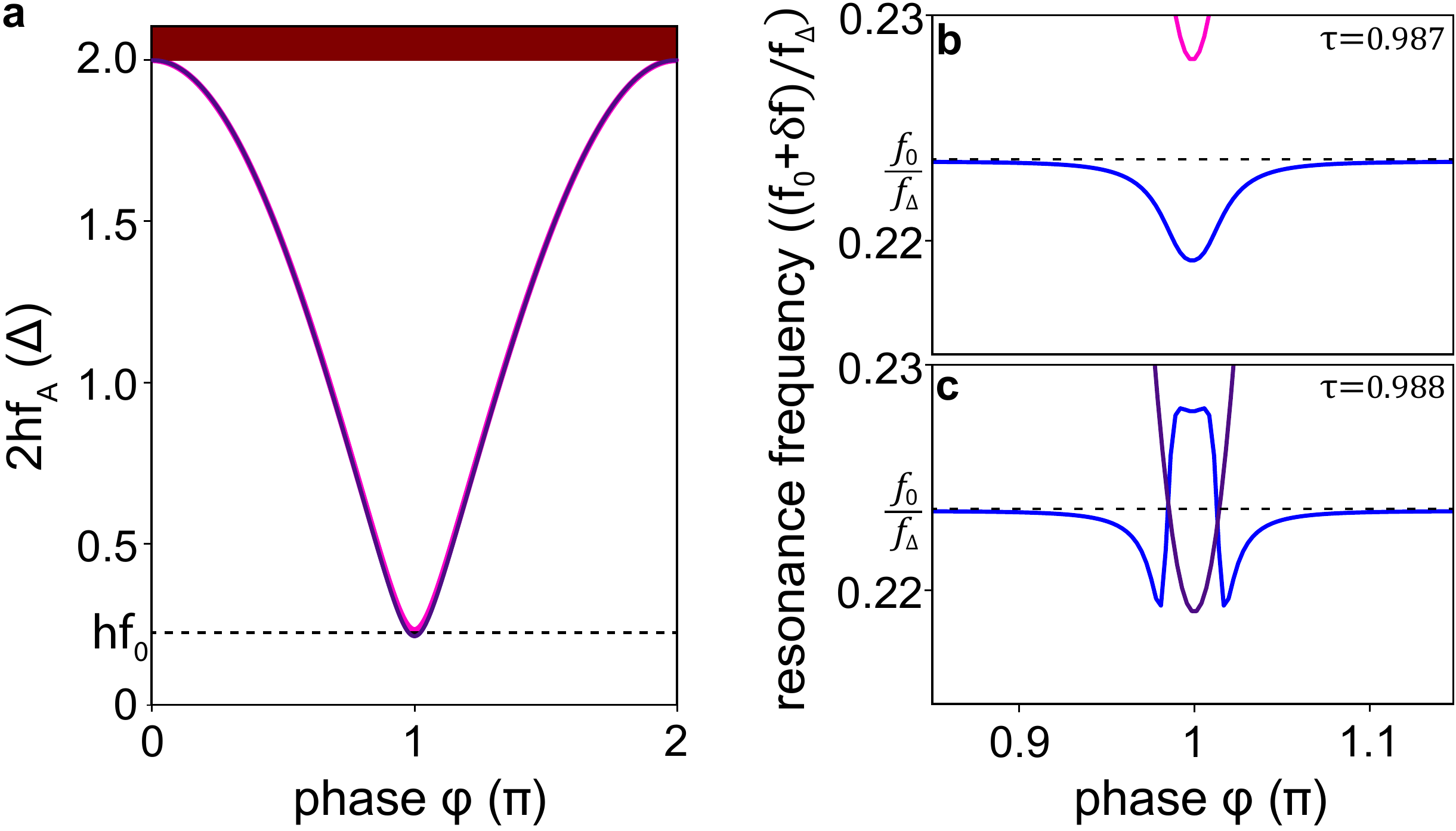}
   \caption{\textbf{Examples of dispersive coupling of cavity to a short channel.} \textbf{a}, Pair transition energies ($2hf_{A}$) for the short junction limit for two different transparencies. Magenta and dark purple traces indicate the quasi-particle transition energy $2hf_{A}$ for transparencies of $\tau=0.987$ and $0.988$, respectively. \textbf{b}, \textbf{c}, Cavity frequency shift $\delta f$ from the bare cavity frequency $f_{0}$ normalized by the frequency corresponding to the gap edge $f_{\Delta}=\Delta/h$ for both $\tau=0.987$ and $0.988$, respectively. The effective coupling strength at $\varphi=\pi$ is $g_{\mathrm{eff}}=0.0139f_{\Delta}$ for both transparencies. As can be seen in \textbf{b}, when the bound states and cavity detuning remains positive the cavity frequency is dispersively shifted downward. If however the detuning is shifted through zero via the phase as in \textbf{c} the cavity frequency exhibits both up and down shifts. Model parameters for \textbf{b} and \textbf{c} are: $\omega_\mathrm{c}=0.2235f_{\Delta}$, $\gamma=0.0067f_{\Delta}$, $\kappa=6.7\time 10^{-6}f_{\Delta}$, $Z=2\times 10^{-5}$. Model details can be found in the methods section.
   }
   \label{fig:input-output}
\end{figure*}

\pz{Before we present data} \ts{of spectroscopic measurements} \pz{it is useful to take a closer look at the predicted dispersive response of the cavity to a single Andreev bound states in the short channel case (cf. dashed line in Fig. \ref{fig:setup}b) \cite{Zazunov03,Tosi19, Hays18}.} Fig.~\ref{fig:input-output}a shows the energies for two-particle transitions with two very slightly different values of transparency $\tau$. Shown in Figs.~\ref{fig:input-output}b and c are the \ts{corresponding} predicted dispersive shifts of the cavity frequency in the weak coupling regime. The response is modelled applying a Jaynes-Cummings type Hamiltonian to an effective two-level system comprised of the ground state and the two quasi-particle excited state of the system, described by Zazunov \textit{et al.} \cite{Zazunov03} for which further details can be found in the Supplementary Information. As can be seen in Fig.~\ref{fig:input-output}b the cavity response is sensitive to the detuning of the two-particle transition to the Andreev bound state from the cavity frequency, $2hf_\mathrm{A}(\varphi)-hf_{0}$. When this detuning is small but positive the cavity resonance is dispersively shifted down in frequency. In contrast if the detuning can be shifted from positive to negative through control of $\varphi$ the resulting dispersive shift of the cavity shifts from downward to upward, as shown in Fig.~\ref{fig:input-output}c. \ts{Thus, from the type of cavity response, information on the detuning can be gained.} Note that in the strong coupling regime, where the effective coupling $g_\mathrm{eff}$ is much greater than the lifetime of photons and Andreev states, such avoided crossing regions would manifest zero-mode Rabi splittings as has been previously reported in both atomic \cite{Janvier15} and nanowire \cite{Hays18} Josephson junctions.

\section*{Spectroscopic measurements}

\begin{figure*}[!t]
   \centering
\includegraphics[width=0.65\textwidth]{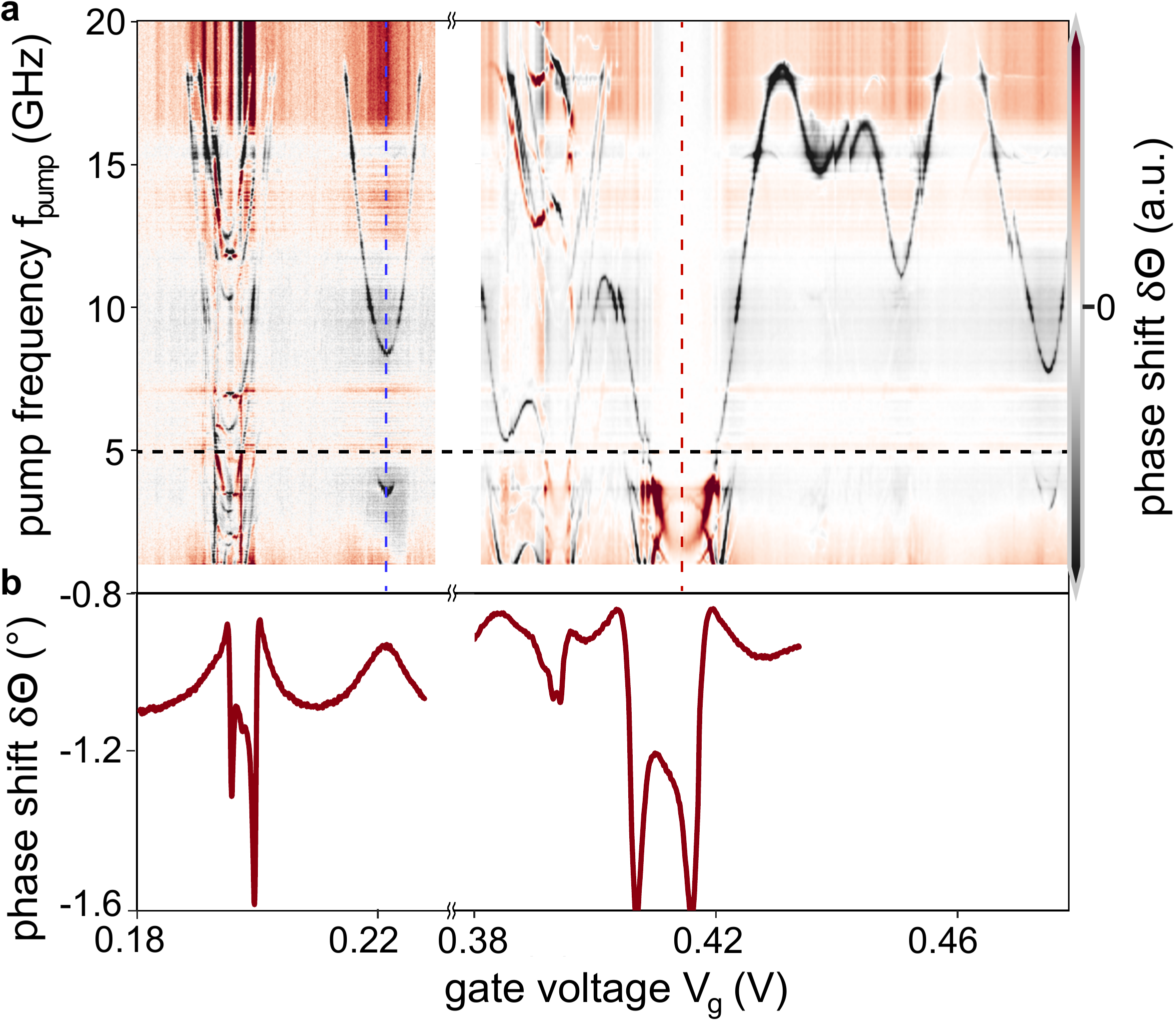}
   \caption{\textbf{Gate-dependent Andreev bound state spectrum.} \textbf{a,} Two-tone spectrum at $\varphi=\pi$ in a gate voltage range from $0.18$ to $0.23\,$V and from $0.37$ to $0.48\,$V. The phase shift of the resonator $\delta\Theta$ is color coded. The probe frequency is indicated by a dashed black line. We assign the abrupt discontinuities in the state dispersion, e.g. around $V_\mathrm{g}=0.43\,$V, to random reconfigurations of the nanowire surface potential or changes in the effective gate charge. The vertical dashed lines indicate the gate voltages chosen for the flux-dependent measurements discussed later. \textbf{b,} Corresponding resonator phase-shift obtained by single-tone spectroscopy in the same gate voltage range. The slight shifts compared to the measurement in \textbf{a} are caused by a time-dependent gate drift.}
   \label{fig:gate-spectrum}
\end{figure*}

After discussing the properties of the $\lambda/4$ resonator coupled to the device chip, we now proceed with spectroscopic measurements on our InAs-nanowire/Al junction. As a first step, gate-dependent experiments were performed at a superconducting phase difference across the junction $\varphi = \pi$ to identify regions with transitions between Andreev levels. Figure~\ref{fig:gate-spectrum}a shows a two-tone spectrum as a function of gate voltage alongside a measurement of the phase shift in the single-tone response with only probe tone applied at the bare cavity frequency $f_{0}$ (cf. Fig.~\ref{fig:gate-spectrum}b).

The phase shift of the single-tone response (shown in Fig.~\ref{fig:gate-spectrum}b) reflects the detuning of the Andreev transitions from the cavity frequency. It reveals dispersive shifts of the cavity resonance due to interaction with the Andreev level(s), which are consistent with the states observed in the two-tone spectrum. The observed modulation of the single-tone response as a function of the applied gate voltage is caused by changes in the transition energy between different Andreev levels. At gate voltages for which the Andreev level transition energy is higher than the resonator photon energy (such as $V_\mathrm{g}\sim 0.22\,$V), the cavity resonance frequency is dispersively shifted down, giving a relative increase in phase. In contrast, at gate voltages for which the detuning between Andreev transition energy and cavity frequency is reduced to zero, an abrupt reversal of the frequency shift is observed, inducing strong negative phase shifts (such as at $V_\mathrm{g}\sim 0.41\,$V). Around zero detuning, despite the signature of an avoided-crossing-like transition, we observe no zero-mode or vacuum Rabi splitting that would indicate strong coupling. The most likely reason are the comparably short lifetimes of the Andreev states being studied, as such the device is being probed in the weak coupling dispersive regime.

The two-tone spectrums in Fig.~\ref{fig:gate-spectrum}a contains richer features than suggested in the single-tone response. For careful selection of a gate voltage such as the region around $V_\mathrm{g}\sim 0.195\,$V or $0.39\,$V the two-tone spectrum reveals transitions to multiple Andreev bound states as well as multiple-photon processes. The multiple photon replicas of the quasiparticle pair transitions to the Andreev states are particularly evident in the low second tone frequencies below the cavity frequency ($<5\,$GHz).

\begin{figure*}[!t]
   \centering
\includegraphics[width=0.85\textwidth]{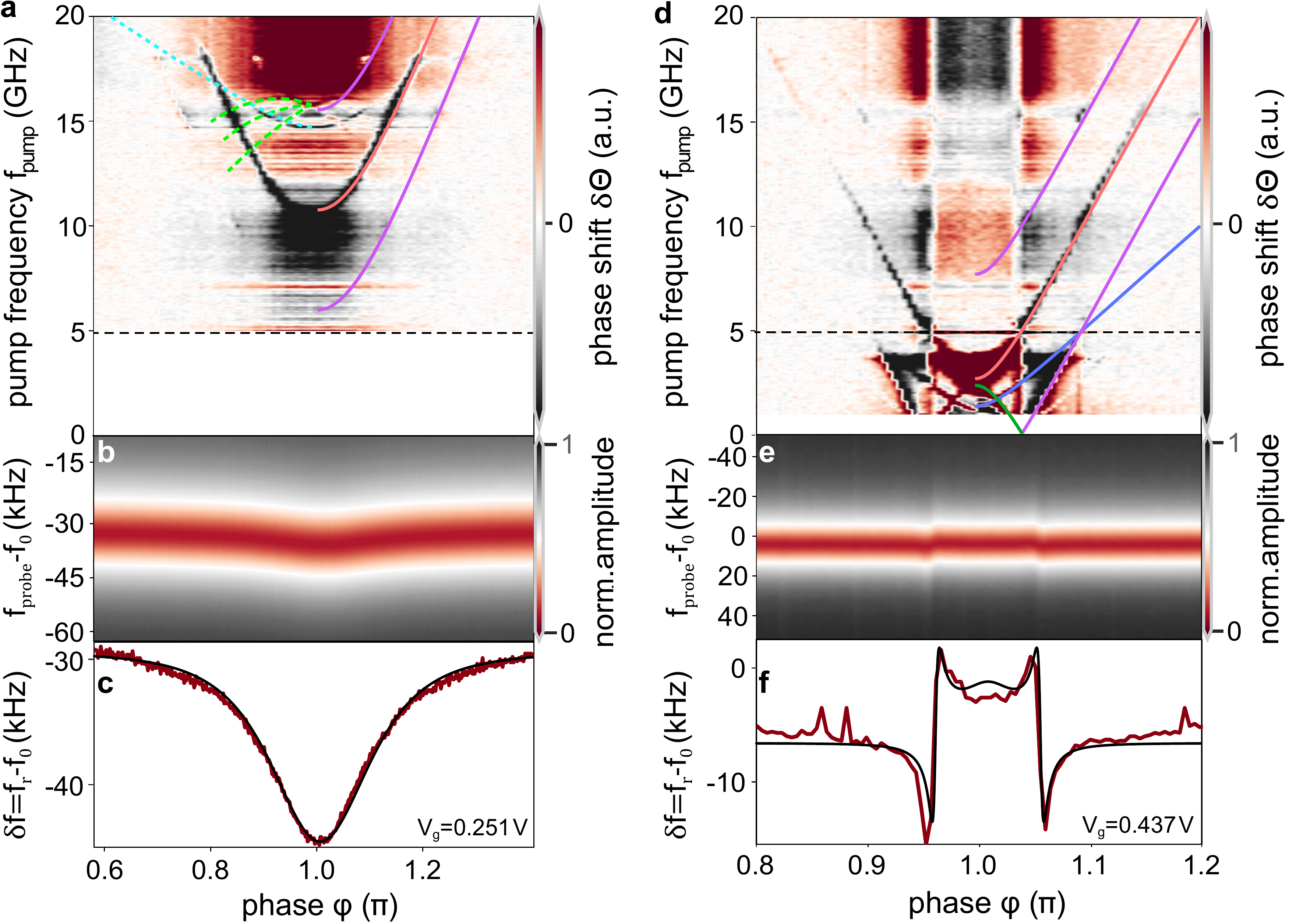}
   \caption{\textbf{Flux-dependent state dispersion.} \textbf{a,} Two-tone Andreev bound state spectrum at a gate voltage of $V_\mathrm{g}=0.251\,$V (blue dashed line in Fig.~\textbf{4$\,$}a) as a function of the magnetic field induced phase difference $\varphi$ across the junction. The red line is a fit to the pair transition driven by the pump tone using the short channel model with parameters $\tau=0.9562$, $\Delta=104\,\mu$eV. The purple lines correspond to a the corresponding two photon process caused by the absorption or emission of an additional resonator photon. The dashed blue line indicates another two-particle transistion feature from another set of Andreev states. Dashed green lines highlight a set of single-quasiparticle transitions between spin polarized Andreev levels. \textbf{b,} Single-tone measurement of cavity response at the same $V_\mathrm{g}$ as in \textbf{a}. The resonance frequency is extracted and plotted in \textbf{c} alongside a fit to a Jaynes-Cummings Hamiltonian (solid black line). Fit parameters are $\gamma=1.32\,$MHz, $Z=6.318\times 10^{-7}$, $f_\mathrm{offset}=-29\,$kHz, $g_\mathrm{eff}(\pi)=60\,$MHz, $\kappa=0.1964\,$MHz. \textbf{d,} Two-tone spectrum at a gate voltage of $V_\mathrm{g}=0.4374\,$V (red dashed line in Fig. ~\textbf{4$\,$}a). Again a prominent two particle transition is observed and fitted with a short channel model with $\tau=0.9981$, $\Delta=132\,\mu$eV (solid red line). Two photon transitions (solid purple lines) which utilize both cavity and pump photons are also observed. Additionally, a solid blue line indicates a two photon pair transition using two pump photons. The solid green line corresponds to a higher-order transition due to the combination of a pump-induced single resonator photon and the relaxation of an Andreev level. \textbf{e}, Single-tone measurement of cavity response at the same $V_\mathrm{g}$ as in \textbf{d}. The resonance frequency is extracted and plotted in \textbf{f} alongside a fit to a Jaynes-Cummings Hamiltonian (solid black line). Fit parameters are $\gamma=466\,$MHz, $Z=4.65\times 10^{-7}$, $f_\mathrm{offset}=-8.6\,$kHz, $g_\mathrm{eff}(\pi)=23\,$MHz, $\kappa=0.1964\,$MHz.
   }
   \label{fig:phase-spectrum}
\end{figure*}

Proceeding to the lower gate voltage section in Fig.~\ref{fig:gate-spectrum}a one finds a rather complex spectrum with an abundance of transitions between $V_\mathrm{g}=0.19$ and $0.20$\,V. In addition, on its right side one observes a single spectral trace which is attributed to a pair transition. Once again, the coupling of the Andreev bound states in the junction to the resonator is reflected in the single-tone measurement shown in Fig.~\ref{fig:gate-spectrum}b. A strong modulation of the phase shift is observed at a gate voltage range from $V_\mathrm{g}=0.19$ to $0.20$\,V, where the transitions found in Fig.~\ref{fig:gate-spectrum}a cross the resonator frequency. In addition, around $V_\mathrm{g}=0.22$\,V where the Andreev level approaches the resonator frequency the cavity phase shift has a smooth maximum.

One finds that at larger gate biases in the presented measurement range ($V_\mathrm{g} = 0.42-0.47$\,V) only a single transition is observed in the pump frequency window. We attribute the single transition found at gate biases $V_\mathrm{g} >0.42$\,V to a quasiparticle pair transition $2E_{A}=hf_\mathrm{pump}$ from the ground state \cite{Tosi19,Metzger21}. The modulation with the gate voltage can be attributed to potential fluctuation in the nanowire channel, e.g. in a chaotic open dot with low charging energy coupled to superconductors by tunnel barriers and a diffusive junction with transparent normal superconducting interfaces, as discussed by Houzet and Skvortsov \cite{Houzet08}. Indeed, in comparable nanowire-based junctions critical current fluctuations were observed, which were correlated to corresponding conductance fluctuations in the normal state \cite{Doh05,Guenel12}. Since ultimately the critical current is determined by the Andreev levels a variation of these levels with gate voltage will directly result in critical current variations. While only a single pair transition is observed in the two-tone spectrum for $V_\mathrm{g} >0.42$\,V we deduce that further high lying transitions must exist at frequencies beyond $f_\mathrm{pump}=20\,$GHz. The spectroscopy as a function of flux (discussed below) reveals some characteristics typical of long Josephson junctions and relatively large frequency shifts from the bare cavity frequency across the measurement flux range are observed.

We next discuss the phase-dependent measurements and use a combination of single and two-tone measurements to evaluate the coupling of the Andreev levels. In Figs.~\ref{fig:phase-spectrum}a and d are shown for which we select gate voltages of $V_\mathrm{g}=0.251\,$V and $0.4347\,$V, respectively. The spectum at $V_\mathrm{g}=0.251\,$V corresponds to a gate condition at which the pair transition frequency remains above the cavity frequency at $\varphi=\pi$. In contrast at $V_\mathrm{g}=0.4347\,$V the pair transition frequency dips below the cavity frequency at $\varphi=\pi$.\footnote{Owing to a gate drift the gate voltage was slightly readjusted between phase-dependent and gate-dependent measurements.} The most prominent feature in both sets of phase data is the pair transition where $2E_\mathrm{A}= h f_\mathrm{pump}$. The pair transition can be fitted using solutions to a transcendental (non-algebraic) expression for the Andreev level energy in a long junction with multiple transverse 1-dimensional normal bands, as discussed in Refs.~\cite{Tosi19,Metzger21} (see Supplementary Information). Features can also be well fit using the short channel analytical solution for the Andreev state energies if the gap $\Delta$ is taken as a free parameter which is reduced below the expected bulk gap (assumed to be $\Delta_\mathrm{Al}\sim 180\,\mu$eV based on previous dc transport studies \cite{Zellekens20a}). We note both fitting methods produce comparable results in the measurement parameters space but differ at higher pump frequencies. In Figs.~\ref{fig:phase-spectrum}a and d fits are performed using the short channel expression shown as solid red lines with parameters as indicated in the caption and results are later used to evaluate the coupling strength assuming coupling with only a single two-level system. The reduced gap required for fitting is a consequence of the states themselves being intermediate between the short channel limit and a long junction (see the Supplementary Information for further discussion). In either case we observe a high transparency $\tau=0.995$ and $0.998$ with fitted $\Delta\sim 104$ and $132\,\mu$eV in Figs.~\ref{fig:phase-spectrum}a and b respectively. We note that fitting with the long junction expression also produces comparable high transparencies (see Supplementary Information).

Apart from the pair transition features in Figs.~\ref{fig:phase-spectrum}a and d, we also find weaker transitions. First, we identify replicas of second order transitions involving absorption or emission of a resonator photon, i.e. at $2E_\mathrm{A} \pm f_\mathrm{res}= h f_\mathrm{pump}$, respectively (solid magenta lines). Furthermore, one finds a transition involving a pair of pump photons, i.e. $2E_\mathrm{A} = 2h f_\mathrm{pump}$ (dashed blue line). The appearance of these replicas is likely related to relatively high power of both pump and probe tones. Interestingly, close to $\varphi = \pi$ we also find a transition with an inverted dispersion (solid green line). We attribute this feature to a higher-order transition where a single resonator photon is created upon pump irradiation with the assistance of the Andreev energy. The relatively prominence of this feature also suggests elevated thermal occupation of the excited states of the Andreev system.

In addition to the two-tone measurement we present the single-tone response in Figs.~\ref{fig:phase-spectrum}b and e as well as the extracted positions of the resonance frequency plotted as shift $\delta f=f_\mathrm{res}-f_{0}$ in Figs.~\ref{fig:phase-spectrum}c and f. Here, the resonances are fitted with a simple Lorentzian lineshape with a linear background subtraction to extract $f_\mathrm{res}$. Note that a negative frequency shift seen here is consistent with a positive phase shift in data plotted in the previous gate spectrums (cf. Fig.~\ref{fig:gate-spectrum}) owing to the features of the resonator. It can be seen that the direction of the resonance frequency shift is related to the detuning of the pair transition frequency and bare cavity frequency, as previously discussed. Most prominent modulation of the resonance occurs when detuning is tuned through zero, as seen in Fig.~\ref{fig:phase-spectrum}f, where the resonance frequency shift changes sign \ts{(cf. Fig.~\ref{fig:input-output}c)}. We fit the extracted frequency shifts assuming that only the single pair transition identified in the two-tone measurement contributes to the cavity response using the known analytical expressions for the coupling in the short channel limit \cite{Zazunov03,Janvier15}. This approach is arguably an oversimplification of our system as other states are inevitably involved as will be later discussed but nevertheless serves as a indication of the lower limit of the coupling and a point of comparison with other studies. Fits to Fig.~\ref{fig:phase-spectrum}e indicate an effective coupling strength at $\varphi=\pi$ of $g_\mathrm{eff}(\pi)=60\,$MHz and a two-level system lifetime $\gamma\sim 1.3\,$MHz. Fits to Fig.~\ref{fig:phase-spectrum}f indicate $g_\mathrm{eff}(\pi)=23\,$MHz and a two-level system lifetime $\gamma\sim 466\,$MHz. We note that fitting includes an offset to the frequency shift independent of $\varphi$ that is $-29\,$kHz and $-8.9\,$kHz in Figs.~\ref{fig:phase-spectrum}e and f, respectively. This shift is likely already caused by the coupling with additional states that are possibly beyond the range of applied pump tone.

Close inspection of the two-tone spectrum in Fig.~\ref{fig:phase-spectrum} reveals a set of lines (dashed green) around $15\,$GHz which can be assigned to single quasiparticle transitions \cite{Park17,Tosi19,Metzger21}. These four-fold transitions originate from modifying the dispersion of the Andreev bound states by Rashba spin-orbit coupling. More precisely, the coupling of lowest Rashba-splitted confinement state in the nanowire to next higher ones results in a difference of Fermi velocities for different spin orientations. This in turn results in an effective spin-splitting of the Andreev bound states. Single quasiparticle transitions between pairs of these Andreev bound states then lead to a four-fold transition pattern in the two-tone spectroscopy, as found in our experiment. Single particle transitions between two subbands of the Andreev spectrum requires a relatively long junction and is thus inconsistent with the pair transition states previously discussed. In addition a weak pair transition like feature with a flatter dispersion (dashed blue line) is observed, again consistent with a junction that is effectively longer. It is found to be challenging to produce accurate fits to the experimentally observed features using expressions for the long junction detailed by Tosi \textit{et al.} \cite{Tosi19,Metzger21} which assumes two bands with different Fermi velocities due to the inflence of spin-orbit interaction characterised by parameter $\Lambda_{j=1,2}=L\Delta/\hbar v_{F,j}$ for band $j$. A rough qualitatively matching is obtained by assuming that the single quasiparticle transitions are in-fact a two photon process involving a cavity and pump photon and an example of such a spectrum is shown in the Supplementary Information. Model parameters include $\tau=0.64$ and $\Lambda_{1}=3.17$, $\Lambda_{2}=2.06$ with $\Delta=180\,\mu$eV. We conclude that these feature arise from a lower transparency and relatively longer channel for which such single particle transitions may become visible owing to the reduced separation in energy between the Andreev subbands. We propose two origins for the coexistence of such long and short junction behavior. It is possible that state are related to two different transverse bands of the nanowire with a significant difference in the Fermi velocity. Alternatively, the states may correspond to channels at different locations on the nanowire surface. The nanowire is half covered with the Al shell such that transport on the nanowire surface that is on the underside (the opposite face from regions with aluminium deposition) may have an effective longer junction length. 

\section*{Conclusion}

We have demonstrated a flip-chip method for the coupling of a rf-SQUID device with a high-$Q$ coplanar waveguide microwave cavity for spectroscopic measurements. The separate processing of the rf-SQUID device chip and resonator chip simplifies fabrication and relaxes requirements, i.e. for the resonator chip the low-loss  substrate material could be chosen to improve the intrinsic quality of the superconducting resonator without taking the risk of degradation or destruction of the Josephson junction on the device chip. Using a demonstrator flip-chip device containing an Al-InAs half-shell nanowire Josephson junction, we were able to show that detailed information about Andreev bound states can be obtained with our approach \ts{using the single- and two-tone response.} The concept presented here, can also be employed for more delicate or complex device circuits, e.g. ones containing topological insulator films, where the thermal budget is rather limited, or circuits with a set of gatemon or Andreev qubits.   

\section*{Methods}

\subsection{Fabrication of the microwave circuit}
The circuit containing the resonator is fabricated onto a $c$-plane sapphire substrate \cite{Megrant12}. The $\lambda/4$ resonator is designed in a coplanar waveguide (CPW) geometry with a bare resonance frequency of $f_{0}\sim 5.4\,$GHz. In order to avoid parasitic effects of residual resist or lift-off induced sidewalls, all superconducting structures have been fabricated out of a globally deposited 50$\,$nm thick sputtered niobium film subsequently patterned with a SF$_6$ dry etching procedure. The chosen resonator design utilizes a single port and is arranged with meanders to occupy half of the available $8 \times 4\,$mm$^2$ sapphire substrate, as shown in Fig.~\ref{fig:Sapphire}a. The remaining half of the substrate provides ample space on which to flip the target device chip while minimizing the overlap of device chip and resonator transmission line. In addition, space for gate connections and an additional microwave line for input of additional signals is provided. In order to stabilize induced vortices in an out-of-plane magnetic field, a global array of $4 \times 4\,\mu$m$^2$ wide flux traps was etched into the ground plane and the termination point of the resonator. In Fig.~\ref{fig:Sapphire}c a detail of the target device chip aligned to the termination point of the $\lambda/4$ coplanar waveguide resonator prepared on the sapphire substrate is shown.

\subsection{Fabrication of the device chip and the nanowire Josephson junction}
The target device chips are fabricated on a Si(100) substrate with high resistivity ($\rho > 100\,$k$\Omega$cm) to reduce substrate-induced losses. Local bottom gates are defined using Ti/Pt ($3/10\,$nm) and are covered with a $3\,$nm$/12\,$nm thick Al$_2$O$_3$ and HfO$_2$ dielectric layer that has been globally deposited by atomic layer deposition. Using a CHF$_3$/O$_2$-based dry etching procedure, the latter is then removed from the areas that are not used for the actual formation or control of the SQUID loop or the nanowire Josephson junction. Single InAs/Al nanowire-based junctions are picked up and positioned on the back-gate using a micromanipulator mounted inside a scanning electron microscope (Hitachi FIB NB-5000). The junctions were fabricated by employing a fully in-situ shadow evaporation scheme, as reported in Ref.~\cite{Zellekens20a}. The InAs nanowire has a diameter of $90\,$nm and a length of $6\,\mu$m. The $20$-nm-thick Al half-shells are separated by $90\,$nm. After the nanowire placement, a superconducting loop of sputtered NbTi with a thickness of $80\,$nm is then fabricated using electron beam lithography and DC magnetron sputtering. In addition to the coupling loop, a ground connection pad is added to ensure a ground reference for the applied gate voltage. A scanning electron micrograph of the device circuit is shown in Fig.~\ref{fig:Sapphire}b.

\subsection{Mounting procedure}
The mounting of the target device chip upon the resonator is achieved using a homemade polymer stamping system. The sapphire resonator is mounted on a section of polymer gel sheet (Hakuto Gel Sheet WF-55-X4-A) stuck to a glass slide. Viewing through the back of the sapphire resonator the operator is able to align the device superconducting loop with the coupling section of the $\lambda/4$ resonator and bring the resonator and device chip into contact. The electrical connection between the two chips has been successfully achieved using two methods. In one a 2-$\mu$m-thick indium pad is added to the gate and ground contact regions of the sapphire chip and the device chip is heated to $\sim 120\,^{\circ}$C when brought into contact with the sapphire. Once the indium is observed to be displaced by the contact of the two chips the temperature is reduced. Alternatively, it was found that a small amount of silver epoxy (either Creative Materials 118-15 or Epo-tek E4110-PFC) at the gate and ground contact regions is sufficient to adhere the two chips and provide good electrical contact. Of the two methods the indium contacts provide a stronger bond but the epoxy method is more forgiving in the event of a mistake in alignment, as this can be corrected before the epoxy is cured by carefully removing the epoxy, cleaning and retrying.

\subsection{Measurement setup}
Spectroscopy measurements were performed in a BlueFors LD400 dilution refrigerator with a base temperature of $14\,$mK. The phase-bias ($\varphi$) across the junction was imposed by threading a small perpendicular magnetic field which is generated by a home-made superconducting coil through the superconducting NbTi loop. The magnetic flux is directly related to the junction phase difference through $\varphi=2\pi\Phi/\Phi_{0}$, where $\Phi_{0}$ is the magnetic flux quantum. The bottom-gate is connected to a bias-tee in order to apply a dc gate-bias $V_\mathrm{g}$ as well as to supply a microwave signal with frequency $f_\mathrm{pump}$ for two-tone spectroscopy. The device is characterized by single-tone as well as by two-tone spectroscopy. For the former the shift of the resonator frequency $f_\mathrm{r}-f_{0}$ induced by the coupling to the weak link is measured by applying a continuous signal to the $\lambda/4$ resonator. The measured bare resonator frequency is $f_{0}=4.911019$\,GHz. Characterisation of the bare resonator at $T\sim 250\,$mK was performed in an Oxford Heliox He$^{3}$ refridgerator. Additional details of the experimental setup can be found in the Supplementary Information.

\subsection{Two-tone spectroscopy}
Two-tone spectroscopy is similar to the concept of dispersive readout in terms of an indirect scanning of the coupled quantum system. A key difference is the way in which the second tone is applied to the device. Instead of the resonator, the impedance-matched gate line is used as a second and independent microwave channel. The measurement scheme is as follows:
First, a frequency-dependent measurement of the cavity response is performed to find the resonance frequency $f_{0}$ of the resonator. The latter is then used as the first tone (probe signal) $f_\mathrm{probe}$=$f_{0}$. In order to determine a range of parameters in which the resonator couples to one or more Andreev bound states, the single-tone response is mapped out with respect to the applied gate voltage and magnetic field. Based on these results, a suitable gate voltage V$_{g}$ or flux induced phase difference $\varphi$ is selected which is then kept constant throughout the measurement. Subsequently, the second tone (pump signal) $f_\mathrm{pump}$ is swept over a wide frequency range (here $\delta hf_\mathrm{pump}\approx \Delta_{\text{Al}}/4$). When the pump frequency is resonante with a transition between Andreev states the system will become excited through the absorption of a photon. This induces a significant alteration of the inductance of the SQUID loop and, in the following, induces a dispersive shift of the resonance frequency which may be detected as a change if phase $\delta\Theta$ measured at the probe frequency. Lastly, the remaining electromagnetic variable, i.e. gate voltage V$_{\text{g}}$ or magnetic flux $\Phi$, is varied to change the energy of the bound state. The advantage of this approach is a very high sensitivity and the possibility to directly map out the energy space of the combined system far below and above the actual cavity resonance frequency. Assuming an undercoupled system, the only limitation is set by the bandwidth of the experimental setup and the wiring in the dilution refrigerator.

\subsection{Fitting cavity frequency shifts}

Fitting of the single-tone response of the cavity is performed using the modeled coupling of a short channel from prior studies\cite{Zazunov03,Janvier15} in which the system is simplified to a Jaynes-Cummings type Hamiltonian. Applying input-output theory we find an expression for the cavity reflection:
\begin{equation*}
A=\frac{-i\kappa}{(\omega_\mathrm{c}-\omega_\mathrm{probe})-\frac{i\kappa}{2}+g_\mathrm{eff}(\varphi)\chi} - 1,
\end{equation*}

\noindent where $\kappa$ is the cavity loss including internal loss and the loss through port 1, $\kappa=\kappa_\mathrm{int}+\kappa_{1}$. Term $\gamma$ is the  decoherence which can include both energy relaxation, poisoning and also pure dephasing rate. Term $\omega_\mathrm{probe}$ is the probe frequency applied to the cavity and $\omega_\mathrm{c}$ is the cavity resonance frequency. Parameter $\chi$ is the susceptibility given as
\begin{equation*}
\chi=\frac{g_\mathrm{eff}(\varphi)}{-((E_{A}/\hbar)-\omega_\mathrm{c})+\frac{i\gamma}{2}}.
\end{equation*}
The effective coupling $g_\mathrm{eff}$ is a function of $\varphi$ and is given as \cite{Zazunov03,Janvier15}
\begin{equation*}
g_\mathrm{eff}(\varphi)=M\frac{1}{\hbar}\sqrt{\frac{\hbar\omega_\mathrm{c}}{2l}}\frac{\Delta}{4\Phi_{0}}\frac{E_{A}(\pi,\tau)}{E_{A}(\varphi,\tau)}\tau\sin(\varphi)\tan(\varphi/2),
\end{equation*}
\noindent where $M$ is the mutual inductance between cavity and SQUID loop and $l$ is the resonator length and $\Phi_{0}=\hbar/2e$. For fitting and simulation the expression is simplified to:

\begin{equation*}
g_\mathrm{eff}(\varphi)=\sqrt{Z}\frac{\Delta}{2}\frac{E_{A}(\pi,\tau)}{E_{A}(\varphi,\tau)}\tau\sin(\varphi)\tan(\varphi/2),
\end{equation*}
\noindent where $Z=(M^{2}/\hbar^{2})(\hbar \omega_\mathrm{c}/2L)(1/2\Phi_{0})$. Further details can be found in the Supplementary Information.


\bibliography{2tonebib}

\section*{Acknowledgements}

The authors thank Yasuhiro Nakamura for fruitful discussions. Support from Dr. Florian Lentz, Dr. Stefan Trellenkamp, Matthias Geitner and Rainer Benczek with the required e-beam lithography and deposition of the superconducting material is gratefully acknowledged. Most of the fabrication has been performed in the Helmholtz Nano Facility. 
This work was partly funded by Deutsche Forschungsgemeinschaft (DFG, German Research Foundation) under Germany’s Excellence Strategy—Cluster of Excellence Matter and Light for Quantum Computing (ML4Q) EXC 2004/1—390534769. This work was supported by JSPS Grant-in-Aid for Scientific Research (B) (No. 19H02548), Grants-in-Aid for Scientific Research (S) (No. 19H05610) and Grant-in-Aid for Scientific Research (A) (No. 19H00867).

\section*{Author contribution}
P.\,P., B.\,B. and M.\,I.\,L. performed the nanowire growth by MBE and the deposition of the superconducting shell. P.\,Z. and R.\,D. fabricated the samples for transport experiments and carried out the low-temperature experiments and the subsequent data analysis. P.\,Z. R.\,D. and T.\,S. wrote the manuscript. All authors contributed to the discussions.
\end{document}


\title{Microwave Spectroscopy of Andreev states in InAs Nanowire-Based Hybrid Junctions using a flip-chip Layout (Supplementary Information)}

\author{Patrick Zellekens$^+$}
\email{p.zellekens@fz-juelich.de}
\affiliation{Peter Gr\"unberg Institut (PGI-9), Forschungszentrum J\"ulich, 52425 J\"ulich, Germany}
\affiliation{JARA-Fundamentals of Future Information Technology, J\"ulich-Aachen Research Alliance, Forschungszentrum J\"ulich and RWTH Aachen University, Germany}
\altaffiliation{Present address: RIKEN Center for Emergent Matter Science and Advanced Device Laboratory, 351-0198 Saitama, Japan}

\author{Russell Deacon$^+$}
\email{russell@riken.jp}
\affiliation{RIKEN Center for Emergent Matter Science and Advanced Device Laboratory, 351-0198 Saitama, Japan}


\author{Pujitha Perla}
\affiliation{Peter Gr\"unberg Institut (PGI-9), Forschungszentrum J\"ulich, 52425 J\"ulich, Germany}
\affiliation{JARA-Fundamentals of Future Information Technology, J\"ulich-Aachen Research Alliance, Forschungszentrum J\"ulich and RWTH Aachen University, Germany}


\author{Detlev Gr\"utzmacher}
\affiliation{Peter Gr\"unberg Institut (PGI-9), Forschungszentrum J\"ulich, 52425 J\"ulich, Germany}
\affiliation{JARA-Fundamentals of Future Information Technology, J\"ulich-Aachen Research Alliance, Forschungszentrum J\"ulich and RWTH Aachen University, Germany}

\author{Mihail Ion Lepsa}
\affiliation{Peter Gr\"unberg Institut (PGI-9), Forschungszentrum J\"ulich, 52425 J\"ulich, Germany}
\affiliation{JARA-Fundamentals of Future Information Technology, J\"ulich-Aachen Research Alliance, Forschungszentrum J\"ulich and RWTH Aachen University, Germany}

\author{Thomas Sch\"apers}
\affiliation{Peter Gr\"unberg Institut (PGI-9), Forschungszentrum J\"ulich, 52425 J\"ulich, Germany}
\affiliation{JARA-Fundamentals of Future Information Technology, J\"ulich-Aachen Research Alliance, Forschungszentrum J\"ulich and RWTH Aachen University, Germany}

\author{Koji Ishibashi}
\affiliation{RIKEN Center for Emergent Matter Science and Advanced Device Laboratory, 351-0198 Saitama, Japan}

\hyphenation{}
\date{\today}


\maketitle
\def\thefootnote{+}\footnotetext{These authors contributed equally to this work}\def\thefootnote{\arabic{footnote}}


\section*{Resonator characterization}

 
In order to determine the characteristic properties of the resonators, namely the internal, coupling and loaded quality factors $Q_{\text{i}}$, $Q_{\text{c}}$ and $Q_{\text{l}}$, the circle fit method presented in \cite{Probst15} is applied to the resonator data. This approach employs the full scattering matrix of the VNA signal instead of only the magnitude or phase data and derives the system properties from the \textit{IQ} plane, i.e. $\Re(\text{S}_{21})$ and $\Im(\text{S}_{21})$, using the fitted reflection coefficient $\Gamma$ 
\begin{align}
\Gamma =\left[ \frac{\Gamma_\mathrm{min}+2i \frac{Q_\mathrm{c}Q_\mathrm{l}Q_\mathrm{i}}{Q_\mathrm{c}Q_\mathrm{l}+Q_\mathrm{c}Q_\mathrm{i}+Q_\mathrm{i}Q_\mathrm{l}}\frac{f-f_0}{f_0}}{1+2i \frac{Q_\mathrm{c}Q_\mathrm{l}Q_\mathrm{i}}{Q_\mathrm{c}Q_\mathrm{l}+Q_\mathrm{c}Q_\mathrm{i}+Q_\mathrm{i}Q_\mathrm{l}}\frac{f-f_0}{f_0}}-1\right]\text{e}^{i\phi}+1,
\end{align}
for an overcoupled resonator as a function of the probe frequency with
\begin{equation}
\Gamma_\mathrm{min}=\frac{Q_\mathrm{l}Q_\mathrm{c}-Q_\mathrm{l}Q_\mathrm{i}+Q_\mathrm{c}Q_\mathrm{i}}{(Q_\mathrm{l}+Q_\mathrm{i})(Q_\mathrm{c}+Q_\mathrm{i})} .
\end{equation} Here, $f$ refers to the applied probe frequency, $f_0$ is the resonance frequency and $\phi$ is a phase offset that describes the tilt of the circle in the \textit{IQ} plane.  The advantage of this non-iterative calculation is its independence of properly defined starting values. In addition, unwanted side effects such as the cable delay, phase noise or the gain and attenuation along the fridge wiring are taken into account during the pre-processing of the scattering data. 

\section*{Measurement setup}

\begin{figure*}[hb]
\centering
\includegraphics[width=0.95\textwidth]{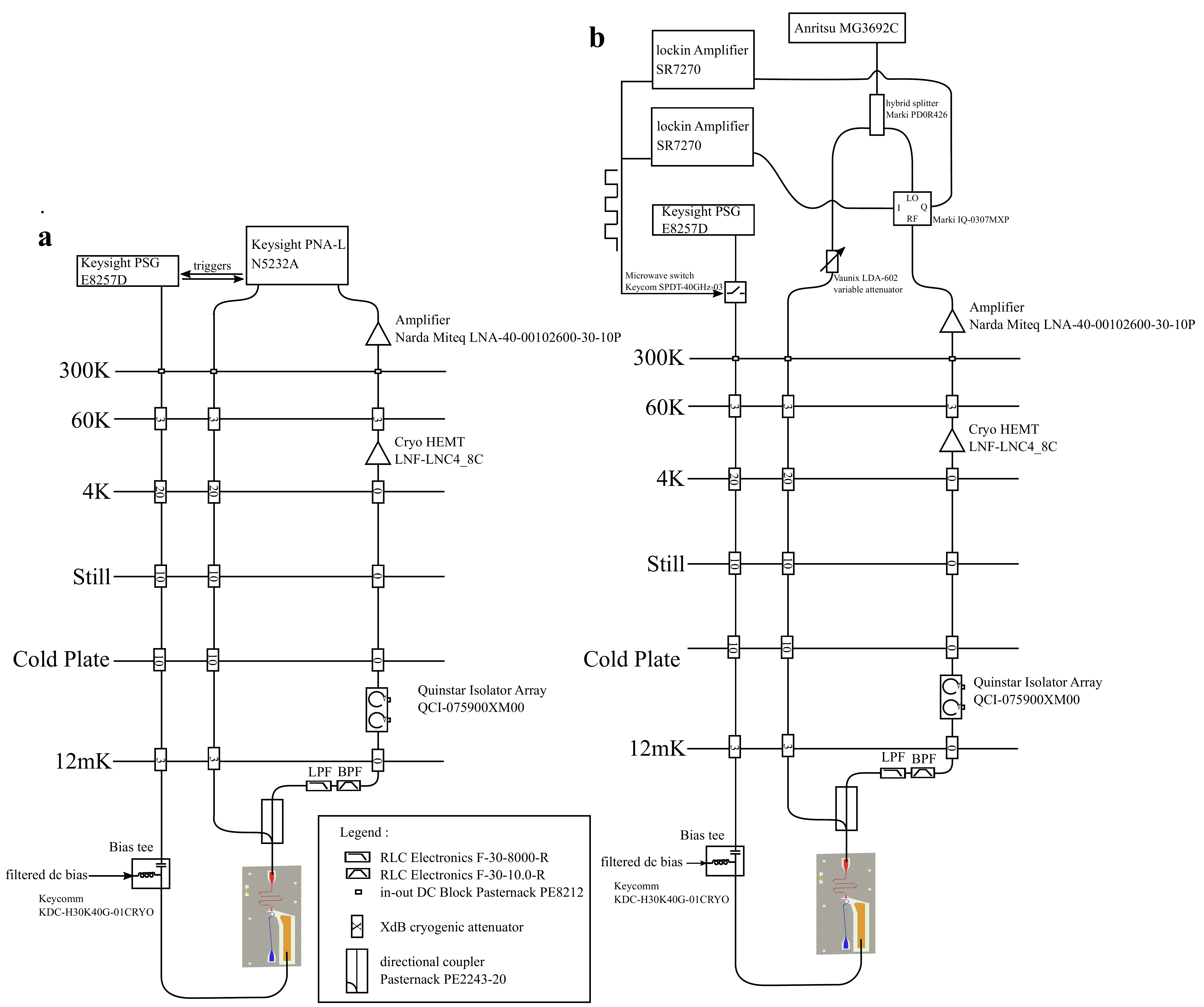}
\caption{\textbf{Schematic of the dilution refridgerator measurement setup}. A simplified schematic of the measurement setup utilized for single-tone and two-tone measurements in \textbf{a} and \textbf{b} respectively. Components ommitted are cold and room temperature microwave switches used for signal routing.}
\label{fig:MeasurementSchematic}
\end{figure*}

Figure \ref{fig:MeasurementSchematic} shows a simplified schematic of the fridge wiring and electronic devices for the two different measurement schemes used within this manuscript. Here, Fig.~\ref{fig:MeasurementSchematic}a corresponds to the vector network analyzer (VNA) driven single tone measurements. The setup depicted in Fig.~\ref{fig:MeasurementSchematic}b, on the other hand, was used for the acquisition of the two-tone spectra. Both layouts have in common that the gate electrode is connected to an off-chip bias tee, thus allowing the simultaneous and independent application of a dc voltage for the electrostatic tuning of the nanowire and a microwave pump tone. Components omitted from the schematic include dc-blocks, room temperature isolators and circulators in the \textit{IQ} mixer circuit and latching microwave switches used to route signals or switch between different measurement setups. Details of filters on the I and Q ports of the mixer are not shown. All semi-rigid fridge lines from the $4\,$K plate down to the mixing chamber plate consist of NbTi. The device itself is protected from stray magnetic fields by means of multiple layers of shielding, including a niobium shield, aluminium shield and an outer cryo-perm shield. In addition, the outer bottom half of the cryostat itself is surrounded by additional mu metal shield. Characterisation of the bare resonator at $T\sim 250\,$mK was performed in a Oxford Heliox He$^{3}$ refridgerator using attenuated superconducting semi-rigid coaxial lines and a vector network analyzer.

\clearpage

\section*{Supplemental - Alignment system}

\begin{figure}[ht]
\centering
\includegraphics[width=0.4\textwidth]{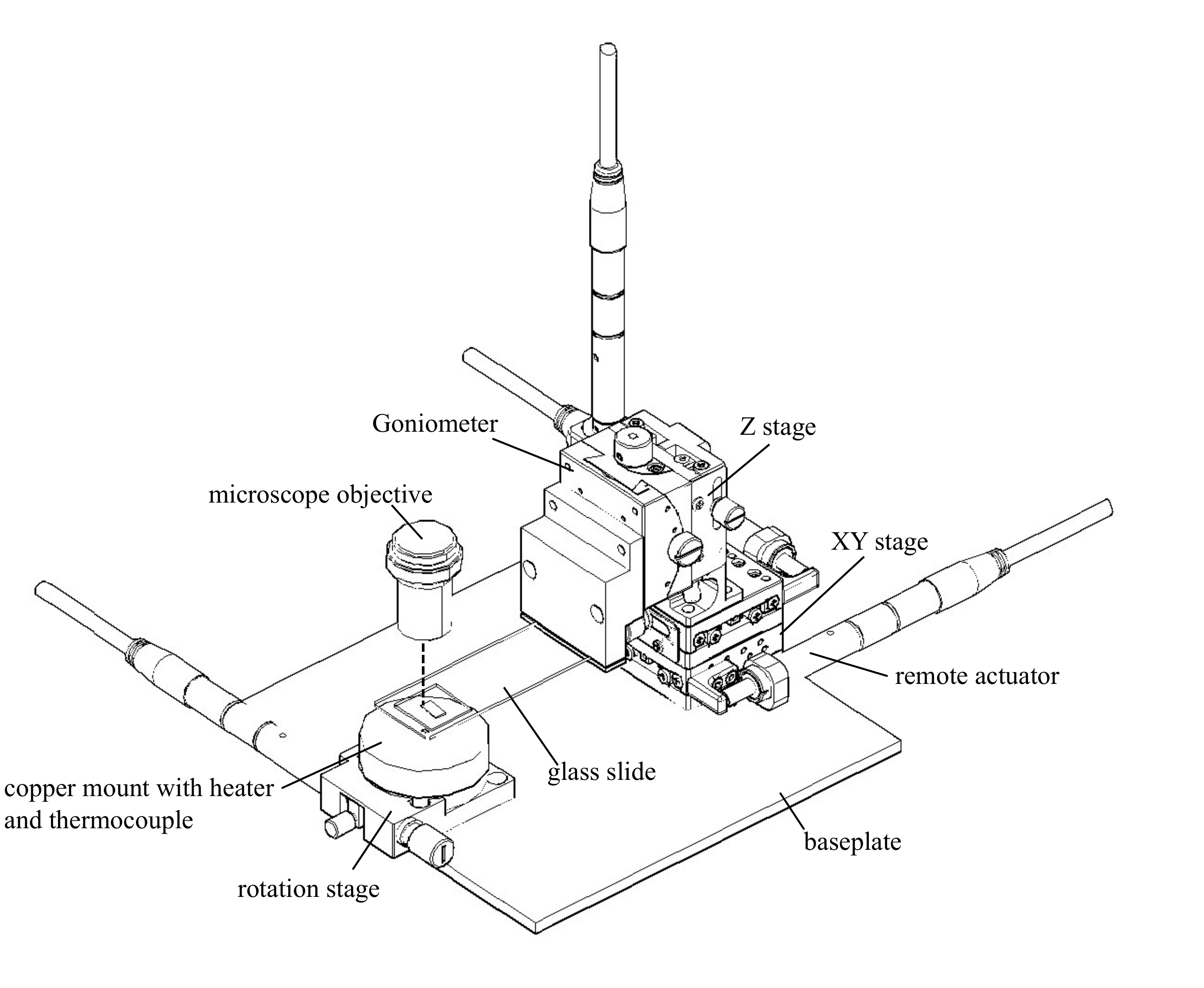}
\includegraphics[width=0.4\textwidth]{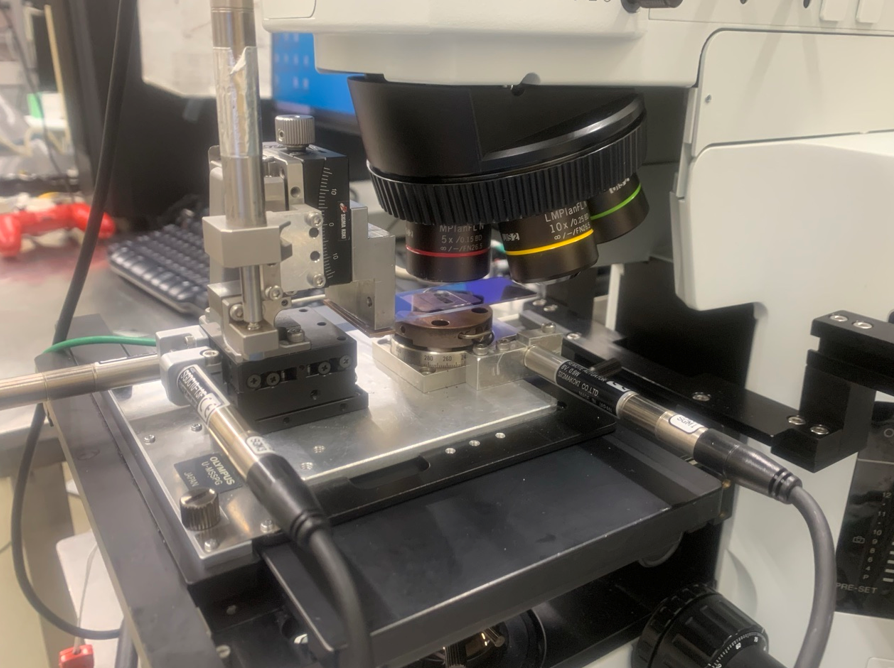}
\caption{\textbf{Schematic of the alignment system.} A schematic of the alignment system alongside an image of the system in use mounted upon the stage of an Olympus BX51 microscope.}
\label{fig:AlignmentSystem}
\end{figure}

The flip-chip processes is performed using a custom made alignment system built from commercial optical stages using the components listed in table~\ref{table:parts}. The system is also employed for stamp transfer of 2D materials \cite{Wang16,Wang17}. The set-up is designed to be mounted on the stage of an Olympus BX-51 microscope modified to extend the clearance between objective and sample stage by $5\,$cm. In addition to commercial components the holder include an aluminium block to grip the glass slide used to hold the resonator chip, an aluminium base plate cut to fit the holder of the microscope stage and a copper mount upon which the nanowire device chip is placed. The copper mount includes thermometry and heaters used to melt the indium or quickly cure epoxy. All components were controlled using a Labview program utilizing a Sony DualShock 4 controller for user input. A schematic of the system and image are shown in Fig.~\ref{fig:AlignmentSystem}.

\begin{table}[!h]
\centering
\begin{tabular}{ |c|c|c| } 
 \hline
 component & device & quantity \\ [1ex]
 \hline
 XY stage & OptoSigma TSDS-402S & x1 \\ 
 Z stage & OptoSigma TSD-403L & x1 \\ 
 rotation stage & OptoSigma KSP-406M & x1\\
 goniometer & OptoSigma GOH-40A15 & x1 \\ 
 remote actuators & OptoSigma SGDC10-13F & x4 \\
 actuator controllers & OptoSigma SRC-101 & x4 \\ [1ex]
 \hline
\end{tabular}
\caption{Commercial components used to assemble the alignment system.}
\label{table:parts}
\end{table}

Two methods have been employed to achieve interconnection of the chips. The first uses indium pads deposited upon the sapphire resonator chips. The second uses small amounts of a silver epoxy (either Creative Materials 118-15 or Epo-tek E4110-PFC) applied to the bonding pads of the device chip applied by transfer using aluminium wire protruding from the capillary of a wedge bonder. In either case the resonator chip is mounted upon a glass slide using a small square of polymer gel sheet (Hakuto Gel Sheet WF-55-X4-A). To prevent device failures due to electrostatic discharge during the process the sections of the glass slide are coated in a film of electron-beam deposited gold and wirebonded to the resonator pads. The nanowire device is placed upon the copper stage and the two parts are brought into contact. The copper stage can then be heated to approximately $150\,^\circ$C to either melt the indium or cure the compressed epoxy and affix the two chips.

\begin{figure*}[hb]
\centering
\includegraphics[width=0.8\textwidth]{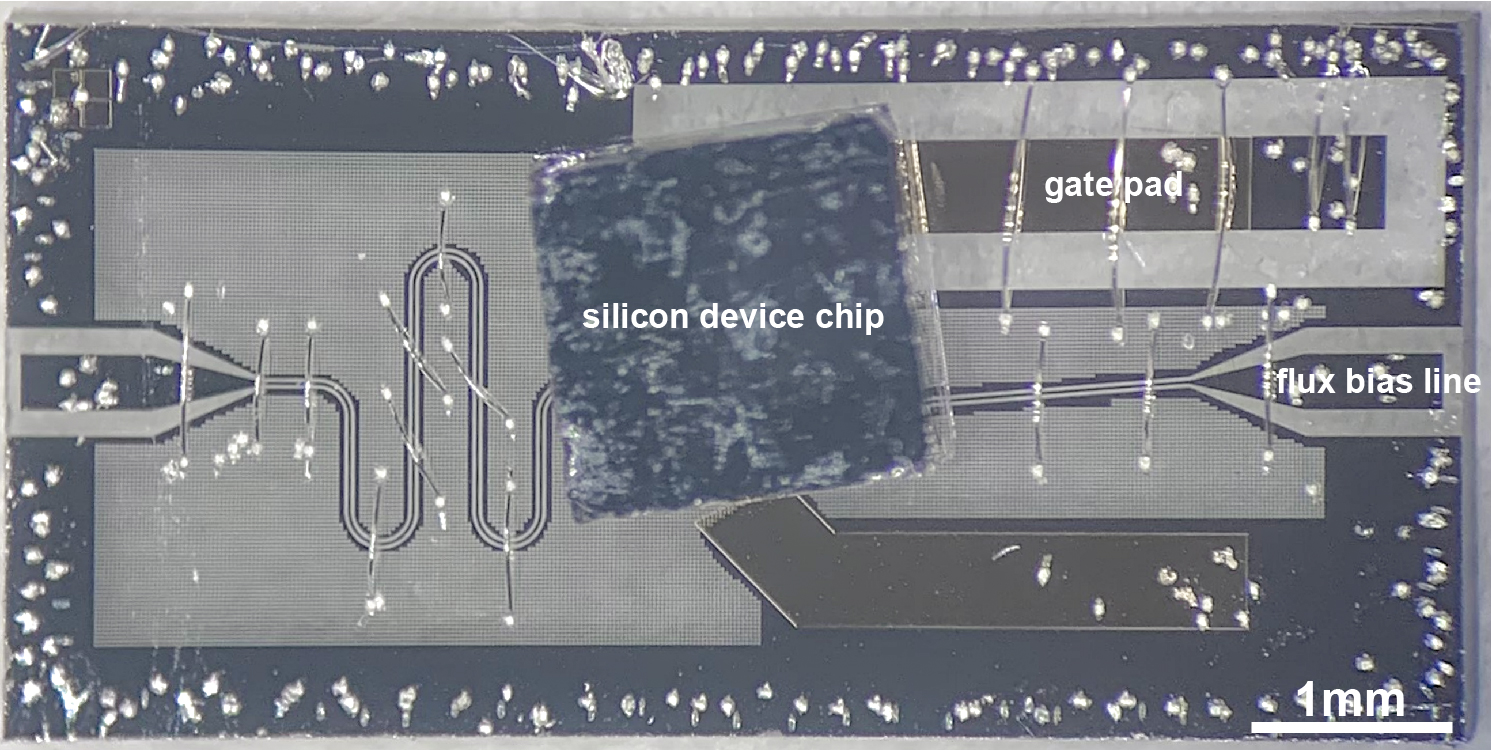}
\caption{\textbf{Example optical image of resonator with flipped device chip mounted.}}
\label{fig:FlippedChip}
\end{figure*}

Fig.~\ref{fig:FlippedChip} shows an example device with flipped chip attached after measurement. Remaining aluminium wire bonds are visible both bridging the resonator coplanar waveguide and around the ground plane perimeter.

\section*{Fitting of phase dependent data with expressions for long/short channels}

In the main text we identify in phase-dependent measurements several features we attribute to two particle transitions to a single Andreev bound state. Here, we discuss the fitting of the two-tone experimental data to known expressions for the eigenenergies of states in both the long and short channel limits. In the short channel limit the Andreev bound states are described by
\begin{equation}
    E_\mathrm{A}(\varphi)=\pm\Delta\sqrt{1-\tau\sin^{2}(\varphi/2)},
\label{eqn:short}
\end{equation}
\noindent where $\tau$ if the transparency and $\Delta$ if the superconducting gap. The Andreev bound states in the short channel limit shifts from the gap edge at $\Delta$ when $\varphi=0$ down to a minimum energy of $\sqrt{1-\tau}$ at $\varphi=\pi$. In the case of the long junction a qualitative form of the Andreev bound state can be captured in a two-band model previously employed by Tosi \textit{et al.} \cite{Park17,Tosi19}. Here they assume transport utilizing two bands within the junction with different Fermi velocities which maybe the result of spin-orbit interaction. The junction of length $L$ includes a barrier at position $x_{0}$. The Andreev bound state energies $E_\mathrm{A}=\epsilon\Delta$ where $\Delta$ is the superconducting gap and $\epsilon$ is a solution to the transcendental equation:

\begin{equation}
    \tau\cos\left[ (\Lambda_{1}-\Lambda_{2})\epsilon \mp \varphi\right] + (1-\tau)\cos\left[ (\Lambda_{1} + \Lambda_{2})\epsilon \frac{2x_{0}}{L}\right] = \cos\left[2\arccos(\epsilon) - (\Lambda_{1} + \Lambda_{2})\epsilon\right],
\label{eqn:transcendental}
\end{equation}

\noindent where $L$ is the junction length and term $x_{0}$ indicates the position of a barrier within the normal region of the junction, $x_{0}\in [-L/2,L/2]$. Terms $\Lambda_{j=1,2}=L\Delta/(\hbar v_{Fj})$, where $v_{Fj}$ is the Fermi velocity for band $j$. 

We fit the observed Andreev bound state features in the two-tone spectrums of the main text using both the short and long channel expressions above. In the case of the long channel we assume two bands associated with the transport have the same Fermi velocity such that $\Lambda_{1}=\Lambda_{2}=L\Delta/(\hbar v_{F})$ and free fitting parameters are $\Lambda_{1}$, $x_{0}$ and $\tau$ with $\Delta=184\,\mu$eV.  In the case of the short channel limit we assign both $\tau$ and $\Delta$ as fitting parameters.

\begin{figure}[ht]
\centering
\includegraphics[width=0.8\textwidth]{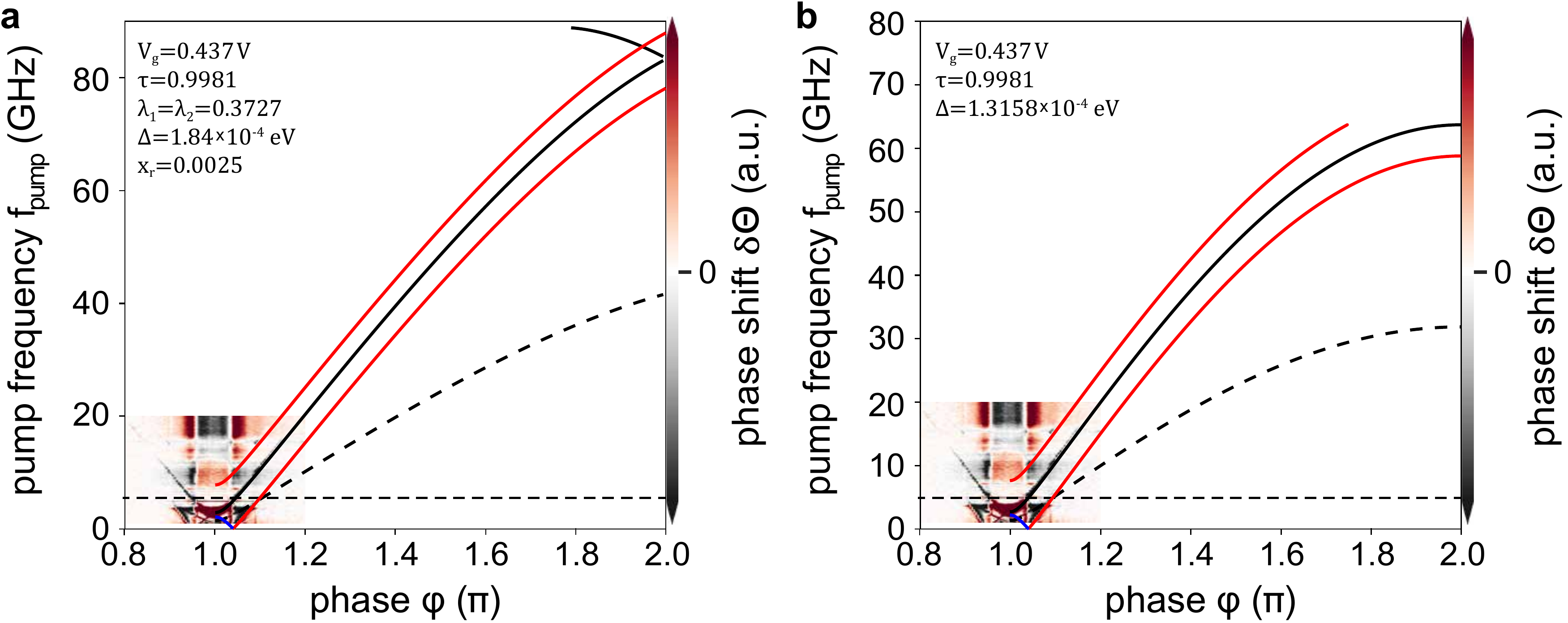}
\caption{\textbf{Example fits to two-tone measurements at $V_\mathrm{g}=437.4\,$mV.} \textbf{a}, Fitting to the long junction expression Eq.~\ref{eqn:transcendental} as solid black lines with the fitting parameters indicated within the figure. The dashed black line indicates the two pump photon process and red lines indicate two-photon processes with both cavity and pump photons. \textbf{b}, Fitting to the short junction Eq.~\ref{eqn:short} as solid black lines with fit parameters, as indicated within the figure. As in \textbf{a} the dashed black line indicates the two-pump-photon process and red lines indicate two-photon processes with both cavity and pump photons.}
\label{fig:FittingData1}
\end{figure}

\begin{figure}[ht]
\centering
\includegraphics[width=0.8\textwidth]{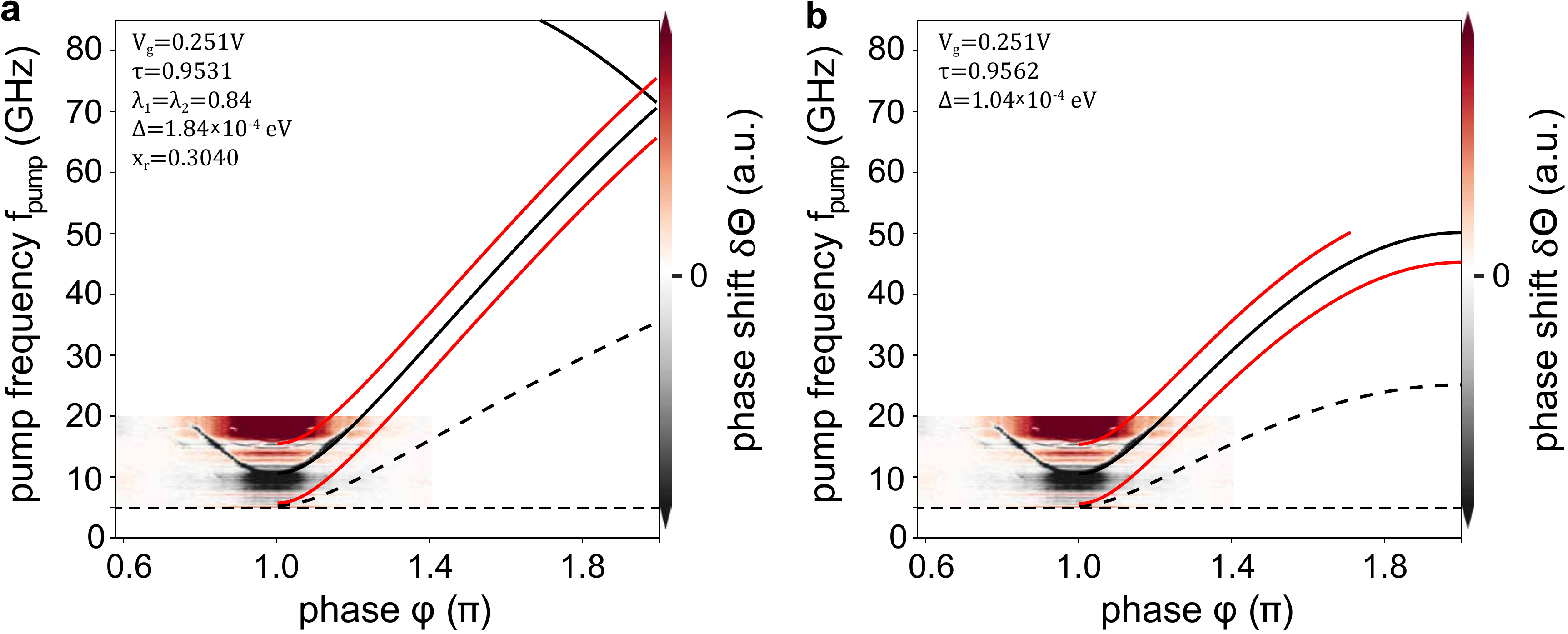}
\caption{\textbf{Example fits to two tone measurements at $V_\mathrm{g}=251.0\,$mV.} \textbf{a}, Fitting to the long junction expression Eq.~\ref{eqn:transcendental} as solid black lines with the fitting parameters indicated within the figure. The dashed black line indicates the two pump photon process and red lines indicate two-photon processes with both cavity and pump photons. \textbf{b}, Fitting to the short junction Eq.~\ref{eqn:short} as solid black lines with fit parameters as indicated within the figure. As in \textbf{a} the dashed black line indicates the two-pump-photon process and red lines indicate two-photon processes with both cavity and pump photons.}
\label{fig:FittingData2}
\end{figure}

We perform fits to both datasets discussed in the main text and present results in Figs.~\ref{fig:FittingData1} and \ref{fig:FittingData1}. We find that fits to the dominant features in the two-tone data can be made with both expressions. Fits using the long junction limit suggest the device is in an intermediate regime between the short limit $\Lambda\ll 1$ and a long junction where the upper bands of the Andreev state are well separated from the gap edge. In this regime it is possible to fit the short junction limit Eq.~\ref{eqn:short} by allowing a reduced superconducting gap to account for the flattening of the dispersion of the state as the system moves away from this limit. The short junction fits are used in the next section with an analytical expression for the coupling of the single Andreev level and the cavity to estimate lower limits for the coupling strength and decoherence times.


\section*{Extraction of parameters for a short channel system}

An estimation of the coupling strength and lifetimes of the system can be achieved using the Jaynes-Cummings type Hamiltonian approach. Here we make the crude assumption that the Andreev bound states are a two-level system (tls) and that only a single tls arising from the short channel is coupled with the resonator. Our evaluation should thus be taken just as an indication of the lower bound for the coupling and lifetime. We therefore consider the regions of gate for which only one dominant interaction with a level is observed within the window of our two-tone measurements. In the short channel limit the Andeev levels have been analytically evaluated as\cite{Beenakker91,Furusaki91,Furusaki99} 

\begin{equation}
    E_\mathrm{A}(\varphi)=\pm\Delta\sqrt{1-\tau\sin^{2}(\varphi/2)}.
\label{eqn:shortABS}
\end{equation}

\begin{figure}[ht]
\centering
\includegraphics[width=1.0\textwidth]{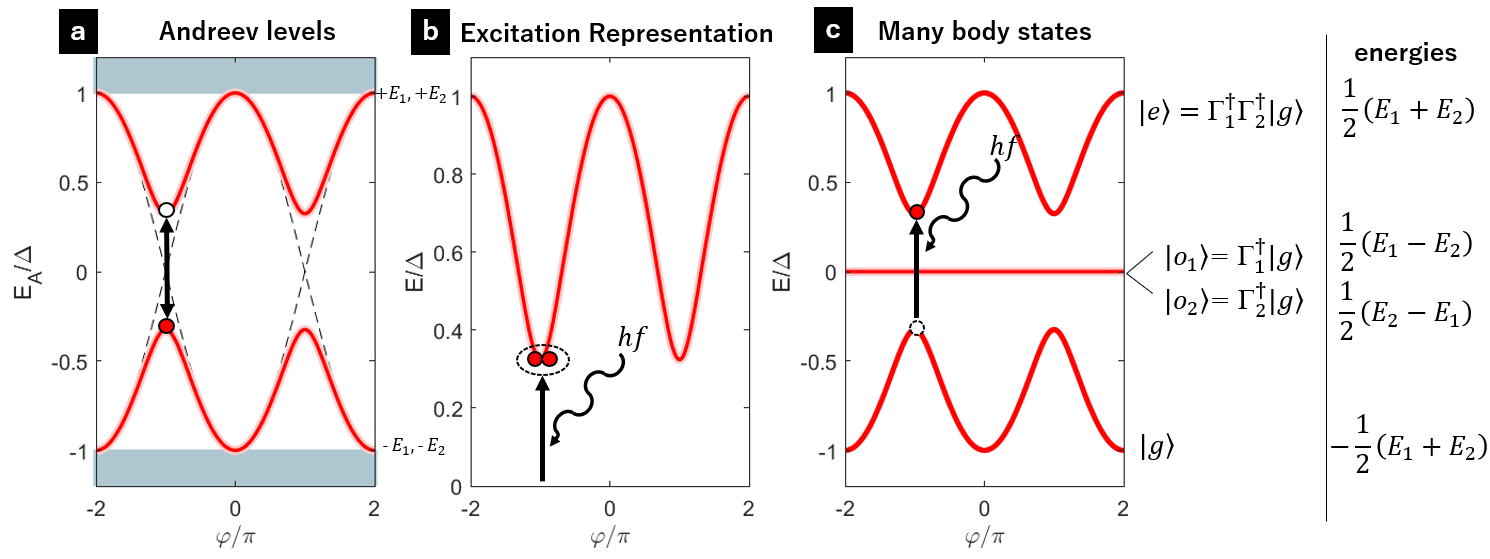}
\caption{\textbf{Representations for states in the short channel limit}. \textbf{a} Andreev levels for the case of $\tau=0.64$. Two sets degenerate levels are at negative and positive energies. \textbf{b} Excitation representation of the coherent two particle excitations which maybe coupled to the cavity photons. \textbf{c} Energies of the many body states of the system showing a single ground $\ket{g}$ and excited state $\ket{e}$ alongside a degenerate pair of odd parity "poisoned" states $\ket{o_{1}}$ and $\ket{o_{2}}$.}
\label{fig:ShortChannelRepresentations}
\end{figure}

For the short channel limit relevant representations of the system are shown in Fig. \ref{fig:ShortChannelRepresentations}. In Fig. \ref{fig:ShortChannelRepresentations} \textbf{a} the Andreev levels of the system are shown, given by solutions to eqn. \ref{eqn:shortABS} for $\tau=0.895$. The Andreev levels are solutions to the de Broglie de Gennes Hamiltonian and come in degenerate pairs ($E_{1}=E_{2}=E_{A}$). Fig. \ref{fig:ShortChannelRepresentations} \textbf{b} shows the excitation representation of the short channel case in which a pair of quasiparticles can be excited with a transition energy of $(E_{1}+E_{2})$. Note that this excitation does not break the pair. In Fig. \ref{fig:ShortChannelRepresentations} \textbf{c} the many body states of the system are displayed which are comprised of combinations of the Andreev levels with energies as indicated. The system ground state $\ket{g}$ and excited state $\ket{e}$ are singlet states with even parity. In addition a doublet of odd parity states ($\ket{o_{1}}$ and $\ket{o_{2}}$) at zero energy are the possible so called "poisoned" states of the system. The relationship between states is indicated using the creation operators $\Gamma_{1} ^{\dagger}$ and $\Gamma_{2}^{\dagger}$ occupying Andreev level $E_{1}$ and $E_{2}$ respectively. The ground and excited state can be considered as a two level system forming the basis for the Andreev level qubit\cite{Zazunov03}.

The system of Andreev tls and cavity can then be described with the following Hamiltonian:


\begin{equation*}
    H = H_\mathrm{cavity} + H_\mathrm{abs} + H_\mathrm{coupling},
\end{equation*}

\noindent consisting of components describing the cavity, Andreev bound states and the coupling between the two. The cavity Hamiltonian is given as
\begin{equation*}
H_\mathrm{cavity}=\hbar\omega_{c}\left(a^{\dagger}a + \frac{1}{2}\right),
\end{equation*}

\noindent where $a^{\dagger}$ and $a$ are the bosonic creation and annihilation operators, $\omega_{c}/2\pi$ is the resonator frequency. In the short channel case with a single bound state the Andreev state Hamiltonian using the Andreev basis is 

\begin{equation*}
H_\mathrm{abs}=-E_\mathrm{A}(\varphi)\hat{\sigma_{z}},
\end{equation*}

\noindent where $E_\mathrm{A}$ is Andreev level energy, and $\hat{\sigma_{z}}$ is the Pauli matrix. The coupling Hamiltonian is related to the inductive coupling between device and cavity and can be expressed as

\begin{equation*}
H_\mathrm{coupling}=M\hat{I}_\mathrm{R}\hat{I}_\mathrm{A}(\varphi),
\end{equation*}

\noindent where $M$ is the mutual inductance, and $\hat{I}_{R}$ is the the transmission line current operator at the device given as
\begin{equation*}
\hat{I}_\mathrm{R}=\sqrt{\frac{\hbar\omega_{c}}{2l}}\left(a^{\dagger}+a\right),
\end{equation*}

\noindent where $l$ is the resonantor length. The Andreev current operator $I_\mathrm{A}(\varphi)$ is a function of the device phase and so does not scale linearly with detuning between cavity and Andreev level transition energy. For the short channel this has been evaluated by Zazunov \textit{et al.} \cite{Zazunov05} as 
\begin{equation*}
\hat{I}_\mathrm{A}=I_\mathrm{A}(\varphi)\left[\hat{\sigma}_{z}+\sqrt(1-\tau)\tan(\varphi/2)\hat{\sigma}_{x}\right],
\end{equation*}

\noindent where $I_\mathrm{A}(\varphi)$ is the Andreev current $I_\mathrm{A}(\varphi)=(2e/\hbar)dE_\mathrm{A}/d\varphi$, and $\tau$ is the junction transparency. The off-diagonal terms of $\hat{I}_{A}$ lead to cavity driven transitions between the ground and excitated states of the system. Writing $\hat{\sigma}_{x}=\hat{\sigma}^{+}+\hat{\sigma}^{-}$, where $\sigma^{+}=\ket{e}\bra{g}$ and $\sigma^{-}=\ket{g}\bra{e}$ the coupling Hamiltonian may be expressed as

\begin{equation*}
H_\mathrm{coupling}=\hbar g_\mathrm{eff}(\varphi)\left(a\hat{\sigma}^{+} + a^{\dagger}\hat{\sigma}^{-}\right), 
\end{equation*}

\noindent such that the Hamiltonian can be reduced to the Jaynes-Cummings model following application of the rotating wave approximation. Here, the effective coupling term $g_\mathrm{eff}(\varphi)$ is given as 

\begin{equation*}
g_\mathrm{eff}(\varphi)=M\frac{1}{\hbar}\sqrt{\frac{\hbar\omega_{c}}{2L}}\frac{\Delta}{4\Phi_{0}}\frac{E_\mathrm{A}(\pi,\tau)}{E_\mathrm{A}(\varphi,\tau)}\tau\sin(\varphi)\tan(\varphi/2),
\end{equation*}

\noindent with $\Phi_{0}=\hbar/2e$. The response of the system maybe further explored using input-output theory using techniques from the study of open quantum systems \cite{Meystre07}. Treating the system as being coupled to a single mode of the cavity with bosonic operator $a$ and possessing a single port (as in the experiment configuration) the equations of motion for $a(t)$ are found as:

\begin{eqnarray*}
\dot{a}=-i(\omega_{c}-\omega_\mathrm{probe})a - \frac{\kappa}{2}a + \sqrt{\kappa_{1}}a_{1,\mathrm{in}}(t) - ig_\mathrm{eff}(\varphi,\tau )\sigma^{-},\\
\dot{\sigma}^{-}=-iE_\mathrm{A}\sigma^{-}-\frac{\gamma}{2}\sigma^{-}-ig_\mathrm{eff}(\varphi,\tau)a,
\end{eqnarray*}

\noindent where $\kappa$ is the cavity loss including internal loss and the loss through port 1, $\kappa_\mathrm{int}+\kappa_{1}$. Here, $\gamma$ is the  decoherence which can include both energy relaxation, poisoning and also pure dephasing rate, while $\omega_\mathrm{probe}$ is the probe frequency applied to the cavity. The input and output of port 1 are given $a_{1,\mathrm{out}}(t)=\sqrt{\kappa_{1}}a(t)-a_{1,\mathrm{in}}(t)$

Taking the stationary limit $\dot{a}=\dot{\sigma}^{-}=0$ the cavity reflection can be evaluated as  

\begin{equation*}
A=\frac{a_{1,\mathrm{out}}}{a_{1,\mathrm{in}}}=\frac{\sqrt{\kappa_{1}}a - a_{1,\mathrm{in}}}{a_{1,\mathrm{in}}}=\frac{-i\kappa}{(\omega_{c}-\omega_\mathrm{probe})-\frac{i\kappa}{2}+g_\mathrm{eff}(\varphi)\chi} - 1,
\end{equation*}

\noindent where $\chi$ is the susceptibility given as

\begin{equation*}
\chi=\frac{g_\mathrm{eff}(\varphi)}{-((E_\mathrm{A}/\hbar)-\omega_{c})+\frac{i\gamma}{2}}.
\end{equation*}

\noindent The cavity frequency shift is then $\delta f=\mathrm{Re}(g_\mathrm{eff}(\varphi)\chi)$ and broadening $\kappa/2\pi=2\mathrm{Im}(g_\mathrm{eff}(\varphi)\chi)$. 

\section*{Fitting signatures of single quasi-particle transitions}

\begin{figure}[ht]
\centering
\includegraphics[width=0.8\textwidth]{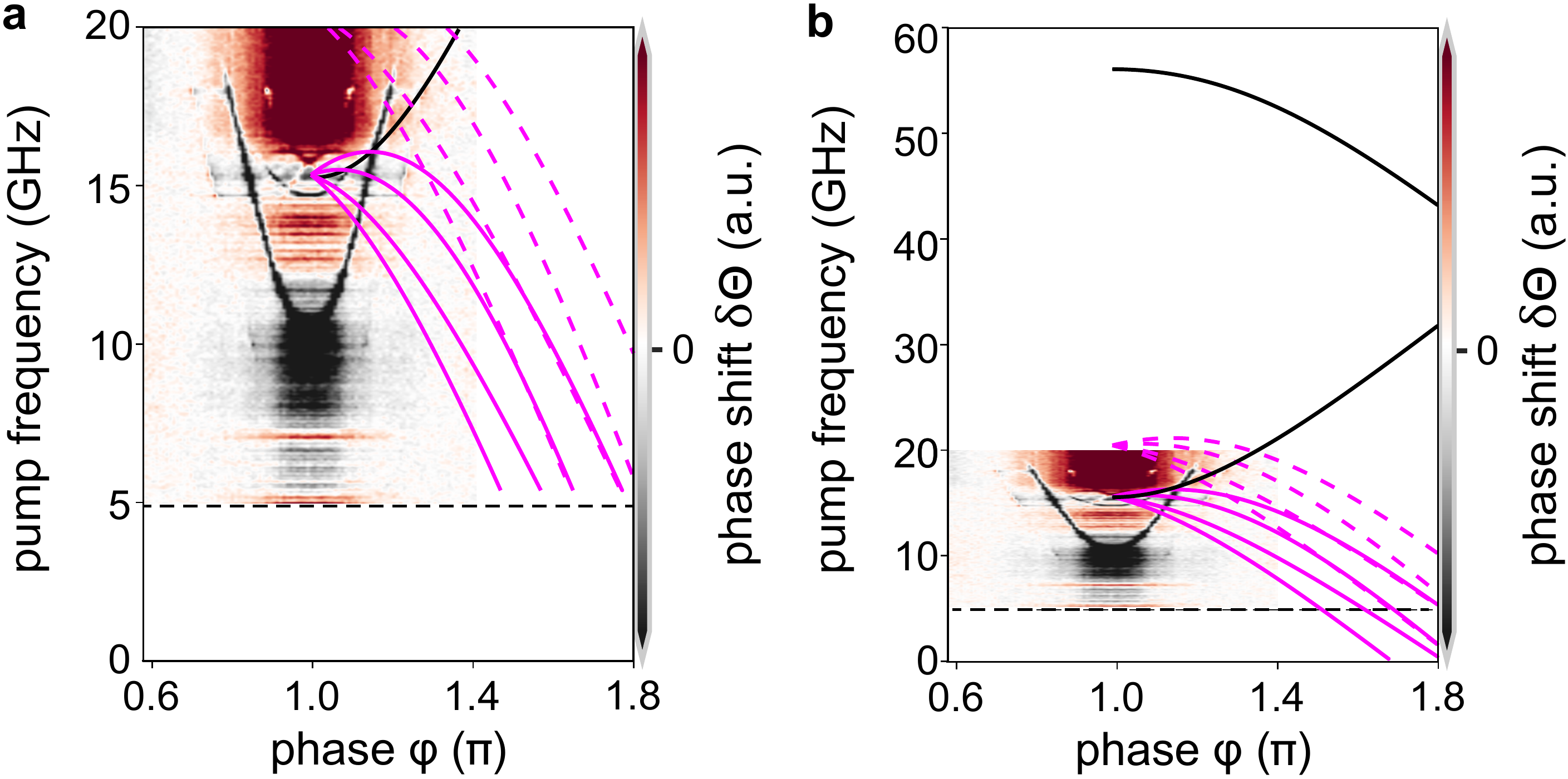}
\caption{\textbf{Example fits to two tone measurements at $V_\mathrm{g}=251.0\,$mV showing fitting to the long junction expression Eq.~\ref{eqn:transcendental}}. \textbf{a} and \textbf{b} show the same fitting result for different range of pump frequency to illustrate the long junction character of the states. Fitting parameters are $\tau=0.64$, $\Lambda_{1}=3.17$, $\Lambda_{2}=2.06$, $x_{0}/L=-0.0002$  and $\Delta=180\,\mu$eV. Solid black lines indicate the two-particle transitions to the lowest subband of Andreev states while dashed magenta lines indicate the single particle transitions between lower and upper subbands. The solid magenta lines then indicate a two-photon process for the same single-particle transitions with absorption of both pump and resonator photons.}
\label{fig:FittingData3_final}
\end{figure}

As discussed in the main text, for some gate voltage we observe features we associate with single quasi-particle transitions between two subbands of Andreev bound states which would only be accessible in the case of $\Lambda>>1$. These features are thus likely related to different states from those discussed earlier that arise from a relatively short channel (or a channel of intermediate length). In Fig.~\ref{fig:FittingData3_final} we attempt a fitting of these features using long junction expression given in Eq.~\ref{eqn:transcendental} with a few assumptions which were found to be necessary to achieve qualitative results comparable to the experimental data most importantly that the feature observed in the detection range associated with the single particle transition is in fact a two-photon feature utilising both cavity and pump photons. The results indicates a relatively long junction with $\Lambda_{1}\sim 2.6$ and $\Lambda_{2}\sim 1.7$ with a comparatively smaller transparency of  $\tau=0.6498$. We speculate that this effective coexisting long junction may arise either from the presence of an additional transverse band with a significantly different Fermi velocity or from an effective longer channel length due to states on the nanowire on the side opposite that upon which the in-situ aluminium contacts are deposited.


\bibliography{2tonebib}